\documentclass[twocolumn,prl]{revtex4-1}

\usepackage{graphicx}
\usepackage{epsfig,psfrag,subfigure}
\usepackage{amsmath, amssymb}
\usepackage{verbatim}
\usepackage{bm}

\newcommand{\gammabar}{\ensuremath\gamma\kern-0.53em-}

\begin{document}

\title{Synthetic Topological Qubits in Conventional Bilayer Quantum Hall Systems}
\author{Maissam Barkeshli and Xiao-Liang Qi}
\affiliation{Department of Physics, Stanford University, Stanford, CA 94305 }

\begin{abstract}
The idea of topological quantum computation is to build powerful and robust quantum computers with certain macroscopic quantum states of matter called topologically ordered states. These systems have degenerate ground states that can be used as robust ``topological qubits" to store and process quantum information. However, a topological qubit has not been realized since the proposed systems either require sophisticated topologically ordered states that are not available yet, or require complicated geometries that are too difficult to realize. In this paper, we propose a new experimental setup which can realize topological qubits in a simple bilayer fractional quantum Hall (FQH) system with proper electric gate configurations. Compared to previous works, our proposal is accessible with current experimental techniques and only involves well-established topological states. Our system can realize a large class of topological qubits, generalizing the Majorana zero modes studied in the recent literature to more computationally powerful possibilities. We propose three tunneling and interferometry experiments to detect the existence and non-local topological properties of the topological qubits.
\end{abstract}

\maketitle

One of the greatest challenges of modern physics is to harness the power of quantum mechanics in order to realize quantum computation.
The fundamental obstacle
is the difficulty of maintaining quantum coherence for long times. Among the various approaches,
topological quantum computation (TQC)\cite{nayak2008} is a proposal which relates the protection of
quantum coherence with fundamental properties of some macroscopic quantum
states of matter, called topologically ordered states\cite{wen04}. These are
gapped states of matter with a certain ground state degeneracy determined solely by the spatial topology of the system.
All degenerate ground states are indistinguishable in all local properties, so that the quantum coherence between
them is robust to environmental noise, for a time that scales exponentially in system size.
In two-dimensions, all low energy excitations of topological states are point-like quasi-particles; for a subset of them,
called non-Abelian states, a set of quasi-particles at fixed positions can also carry a robust,
topological degeneracy of states.

The topological degeneracy in both ground states and non-Abelian quasi-particle states can be used as a ``topological qubit," to carry quantum information.
Most proposals use the latter since it is considered easier to realize quantum operations by manipulating the quasiparticle positions.
While the non-abelian states are not yet experimentally established, currently there are two
types of candidate systems for non-Abelian states: The non-Abelian quantum Hall states\cite{nayak2008},
the most promising of which is proposed to be realized at filling $\nu = 5/2$ \cite{moore1991,willett2009}, and topological superconductors
with Majorana zero modes\cite{alicea2012review,mourik2012,rokhinson2012}.
The topological superconductor proposals have the advantage of possibly larger energy gaps
and well-controlled location of non-Abelian defects.
However, the non-abelian defects that can be realized in these proposals are restricted to Ising anyons, which have limited
computational power \cite{nayak2008}, and the required
proximity between superconductor and semiconductor or topological insulator adds complications to probing the non-Abelian anyons.

\begin{figure}
\subfigure[]{
\includegraphics[width=3.3in]{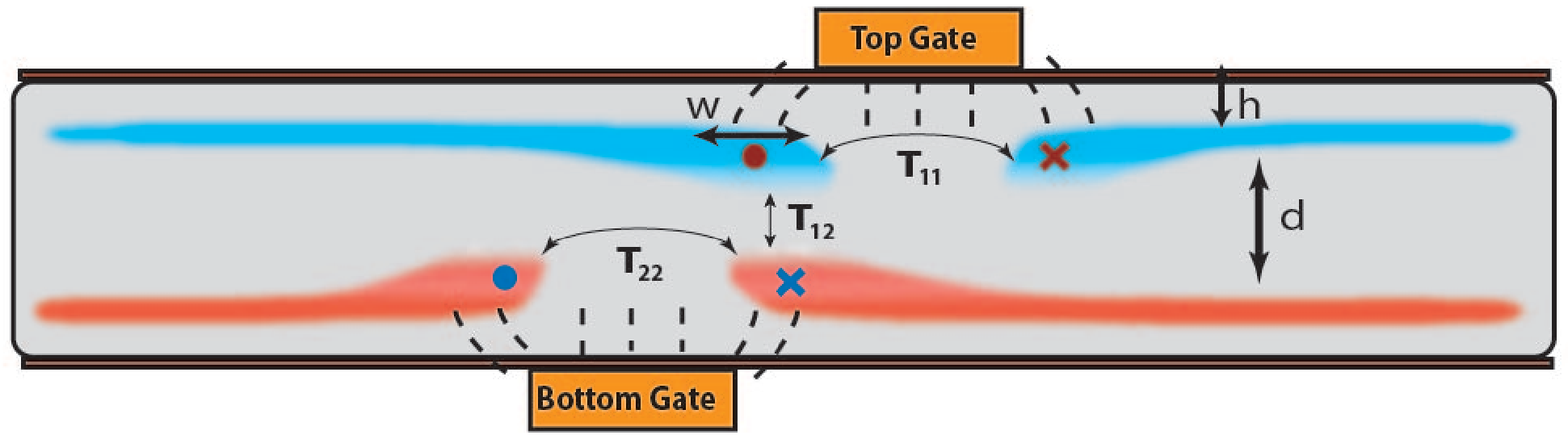}
}
\subfigure[]{
\includegraphics[width=3.5in]{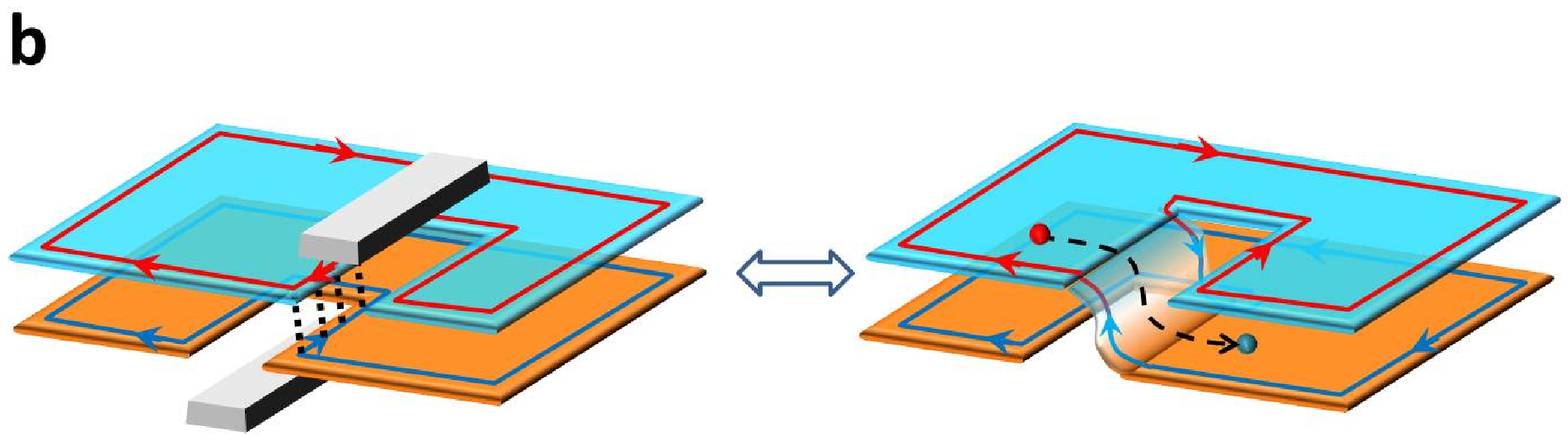}
}
\caption{\label{expsetup}
(a) Cross-section of proposed device.
Top and bottom gates partially deplete the electron fluids,
and the fringe fields cause the wave function at the edges to extend further vertically,
enhancing the tunneling. Filled circles and crosses indicate the gapless chiral edge states, moving out of or into the page.
(b) 3D view. Interlayer backscattering between edge states smoothly connects the layers,
allowing for coherent quasiparticle propagation between layers. The red and blue balls connected by a dashed line illustrates an inter-layer path of the quasiparticle.
}
\end{figure}

In this paper, we propose a new experimental platform that realizes a wide class of non-abelian
defects \cite{barkeshli2010,barkeshli2011orb,barkeshli2010twist,barkeshli2012a,barkeshli2012,you2012,clarke2012,lindner2012,cheng2012,vaezi2012,hastings2012,oreg2013},
vastly generalizing Majorana zero modes,
using conventional Abelian bilayer quantum Hall states. Our proposal is inspired by recent progress in vertical field effect transistors,\cite{sciambi2010,sciambi2011}
and recent theoretical developments regarding extrinsic defects in FQH states.\cite{barkeshli2010,barkeshli2012a,barkeshli2012}
As is illustrated in Fig. \ref{expsetup}, our setup only requires the simplest bilayer FQH states that have been observed in the laboratory \cite{boebinger1990,suen1991,eisenstein1992,lay1997,manoharan1997}, with certain configurations of top and bottom gates. The key idea is to use the gate configuration to induce inter-layer tunneling and thus build ``staircases" that coherently connect the two layers. The end points of such ``staircase" lines are non-Abelian defects,
which we refer to as {\it genons}, although the bilayer FQH state itself is Abelian. The non-Abelian statistics of genons are determined by the parent state of the bilayer, which include the Ising anyons realized in topological superconductors as a special case. With a larger gap than known non-Abelian FQH states and well-controlled position of non-Abelian defects, but without requiring superconductivity, our proposal combines the advantage of both types of proposed systems in the literature.

In what follows, we explain the proposed experimental setup and discuss two
novel experiments that can be used to detect the topological qubit.

\noindent\bf{Experimental Setup} \rm

Our experimental setup (Fig. \ref{expsetup}) consists of a double-layer quantum well system,
with line junctions in each layer that are offset in the lateral direction. One way to create the line junctions
is with top and bottom gates; the gates are used to deplete the layer closest to them,
and to increase the interlayer tunneling over a small region in their vicinity.
Another way to create the line junctions, which has been employed successfully in single-layer systems,\cite{kang2000}
is through a physical obstruction within each layer, such as an undoped Al$_x$Ga$_{1-x}$As/AsAs barrier.
When the layers are well-separated, each layer can individually form its own single-layer FQH state, such as a
filling fraction $\nu = 1/3$ Laughlin FQH state. With larger interlayer interactions,
more exotic interlayer correlated states are also possible. Here we consider the simplest
case where each layer individually forms a $\nu = 1/m$ Laughlin FQH state. Double layer 1/3-Laughlin
states have routinely been realized experimentally, with gaps on the order of several K \cite{boebinger1990,suen1991,eisenstein1992,lay1997,manoharan1997}.

At the boundary of the line junctions, there are gapless chiral edge states. With the offset between top
and bottom gates, and in a suitable parameter region that will be analyzed below, it is possible to induce backscattering
between counter-propagating edge states from \it different \rm layers. The back-scattering can open a gap at the boundary and
effectively build a ``staircase" that coherently connects the two layers (Fig. \ref{expsetup} (b) ).
Such staircases are the essential elements of our proposal, which enable the realization of non-planar
geometries. As will be
discussed below, the end points of a staircase line junction
are non-Abelian topological defects, a pair of which carries topological degeneracy and thus acts as a topological qubit. In the
following we will refer to such a staircase as a twisted line junction (TLJ).

\noindent\bf{Conditions for building TLJ} \rm

The Hamiltonian of the edge states around the line junctions is described  by the chiral Luttinger liquid theory \cite{wen1992},
$H=\int dx\left(\mathcal{H}_0+\delta\mathcal{H}_t\right)$, with
\begin{align}
\mathcal{H}_0 &= m \pi v_0 \sum_{\alpha,\beta, I,J}n_{\alpha I} \lambda_{IJ}^{\alpha \beta} n_{\beta J} ,\label{H0}\\
\delta \mathcal{H}_{t} 
&= \sum_{I,J}2|T_{IJ}| \cos( m(\phi_{LI} + \phi_{RJ}) + \theta_{IJ}).\label{Ht}
\end{align}
$\mathcal{H}_0$ and $\delta\mathcal{H}_t$ are the Hamiltonian densities of the kinetic/interaction terms and the backscattering terms, respectively.
$\alpha,\beta=L,R$ and $I,J=1,2$ denote the chirality and the layer index of the edge states, respectively.
The densities $n_{\alpha I} = \frac{1}{2\pi} \partial_x \phi_{\alpha I}$, with $\phi_{\alpha I}$ the chiral boson fields satisfying the
commutation relations 
$[\phi_{R/L I}(x), \phi_{R/L J}(y)] = \pm i \frac{\pi}{m} \delta_{IJ} sgn(x - y)$. $T_{IJ} = |T_{IJ}| e^{i\theta_{IJ}}$ are back-scattering amplitudes between counter-propagating edge states $LI$ and $RJ$ of the two layers (see Fig. \ref{expsetup}). The velocity of the edge modes, $v_0$, is assumed for simplicity to be the same for all of the edge states. The diagonal entries of $\lambda$, $\lambda_{II}^{\alpha \alpha}$ are normalized to $1$, while the off-diagonal elements of the symmetric matrix $\lambda$ parameterize the density-density interaction between the different edge channels.

To realize the TLJ, the following conditions are required:

1) {\it The line junction of one layer must not destroy the FQH state directly beneath (or above)
it.} In the setup of Fig. \ref{expsetup}, this means the gates must deplete the mobile electron density in
the layer closest to them, while keeping the density in the further layer such that it remains in its FQH plateau.
Based on previous experimental studies,
such a situation is easily possible\cite{sciambi2010,goldhaber}.

2) {\it $T_{12}\gg T_{11},~T_{22},~T_{21}$, {\it i.e.} the dominate back-scattering term should be $T_{12}$. } As long as
the width of the line junction $l$ 
is significantly larger than the inter-layer distance
$d$, this condition can be easily satisfied. If the system is clean, the intra-layer tunneling has a
phase factor $T_{II}\propto e^{ilx/l_B^2}$ (with $l_B=\sqrt{\hbar c/eB}$ the magnetic length) due to the momentum
mismatch, while the vertical tunneling has no momentum mismatch. This further suppresses other back-scattering terms.
Furthermore, near the gates where the density is not depleted, the wave functions will extend further near the other layer,
because the fringe fields of the gate raise the potential well that vertically confines the 2DEG.
This exponentially increases $T_{12}$, as has been observed experimentally\cite{sciambi2010,sciambi2011}.

 3) {\it  The back-scattering $T_{12}$ should generate an energy gap.} In the absence of interactions between edge states,
the tunnelings $T_{IJ}$ are irrelevant perturbations in the renormalization group (RG) sense. In this case, a large tunneling
$T_{12}\gtrsim \hbar v_0/l_B$ is required to open an energy gap and lead to the required TLJ.
However, it is possible for the screened Coulomb interaction between edge states to make $T_{12}$ a relevant
perturbation in the RG sense, so that any small $T_{12}$ will open a gap and allow coherent quasi-particle tunneling across the junction.
    The Coulomb interaction gives an estimate of $\lambda_{IJ}^{\alpha \beta}$: $m \pi \hbar v_0 \lambda_{IJ}^{\alpha \beta} / l_B \sim U_{IJ}^{\alpha \beta}$,
where $U_{IJ}^{\alpha \beta}$ is the Coulomb energy between edges.
When the gates are wide $(l \gg d)$, the only appreciable interaction will be between the two edges on top of each other, which, ignoring
screening due to the gates, is estimated by $m \pi \hbar v_0 \lambda^{RL}_{12} / l_B  \sim e^2/\epsilon d$.
In this case, the scaling dimension of the electron back-scattering term is $\Delta_t = m \sqrt{\frac{1 - \lambda^{RL}_{12}}{1+\lambda^{RL}_{12}}}$ \cite{kane1997}. The tunneling is relevant in the RG sense if $\Delta_t < 2$ in the clean limit, or $\Delta_t < 3/2$ for disorder induced backscattering (assuming Gaussian-correlated disorder). For realistic parameters in GaAs samples, $v_0 \sim 10^5 m/s$, $d_{12}^{RL} \sim 100 nm$, $l_B \sim 20 nm$, $\epsilon \sim 13 \epsilon_0$, we see that the repulsive interaction can easily make $\lambda^{RL}_{12}$ to be order one, so that the tunneling will be relevant.

4) {\it The edges of the line junctions should be sharp enough in order for the edge theory description (\ref{H0}), (\ref{Ht}) to be applicable. } The width of the edge region $w$ (Fig. \ref{expsetup} (a)), in which the density of electrons gradually decreases to zero, must be on the order of $l_B$, so that it can be treated as being part of the gapless edge of the bulk FQH state. 
 $w$ is set by the fringe fields of the gates, and is on the order of the vertical distance $h$ between the 2DEG and the gate, so it is preferable to minimize $h$.
For this reason, in addition to GaAs double quantum wells, two layers of graphene, separated by a thin insulating barrier may be an advantageous system, since $h$
can be made dramatically smaller \cite{sciambi2011}.

Now that we have demonstrated the feasibility of building the TLJ, we will discuss the realization of
topological qubits and experimental predictions.

\bf{Topological degeneracy: Geometrical picture} \rm

A $\nu = 1/m$ Laughlin FQH state on a genus $g$ surface has a ground state degeneracy $m^g$ \cite{wen1990b}.
This can be understood on a torus to be a consequence of the two independent non-contractible
loops, $a$ and $b$,  which intersect only once, combined with the fractional statistics of the
quasiparticles. Together, these imply the following 'magnetic' algebra for quasiparticle loop
operators $\hat{W}(a)$ and  $\hat{W}(b)$: $\hat{W}(a) \hat{W}(b) = e^{2\pi i/m} \hat{W}(b) \hat{W}(a)$.
$\hat{W}(a,b)$ measures the fractional charge inside the loop $a,b$ respectively, which
 can be shown to commute with the low energy Hamiltonian in the ground state
subspace. Therefore, the ground states form an irreducible representation of this algebra, which is at least $m$-dimensional.
A genus $g$ surface gives $g$ independent copies of the magnetic algebra, and therefore a topological degeneracy of $m^g$.
On a finite size system, quasiparticle-quasihole tunneling around $a$, $b$ mixes the different states,
splitting the degeneracy by an energy that is at most exponentially small in system size.

\begin{figure}
\includegraphics[width=3.3in]{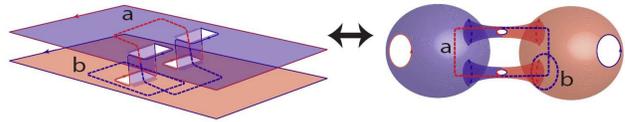}
\caption{\label{doubleStaircase}A pair of TLJs introduces two non-contractible loops,
so the space is topologically equivalent to a torus. The edge states around the outer edge of
the sample and around the gates are mapped to edge states around four holes on the
surface of the torus.
}
\end{figure}

Now consider our experimental proposal (Fig. \ref{expsetup}). With two TLJ's (Fig. \ref{doubleStaircase}), 
the system now has two independent non-contractible loops $a$ and $b$ that intersect only once;
the system is topologically equivalent to a torus with four holes cut out on the surface (Fig. \ref{doubleStaircase}).
With $n$ TLJ's, the system is topologically equivalent to a genus $g = n - 1$ surface, with
$2n+2$ holes cut out on the surface, which will have a topological degeneracy of $m^g$. The
holes cut on the surface contribute chiral edge states with a finite size gap inversely proportional
to the perimeter of the hole, but they do not affect the topological degeneracies at all.
We note that the gapless edge states surrounding the TLJs can also be removed through more
complicated interdigitated gate configurations (see Supplemental Materials).

Since each TLJ increases the degeneracy by a factor of $m$, each end point of the TLJ, viewed as a point defect,
can be associated with $\sqrt{m}$ degrees of freedom. Similar to the Majorana zero mode carrying
$\sqrt{2}$ degrees of freedom, each end point of a TLJ is a non-Abelian topological defect which
we refer to as a {\it genon}.\cite{barkeshli2012} The resulting topological degeneracies can
be used to realize a topological qubit; this setup enables manipulations of such topological qubits
by braiding the genons. The braiding process can be shown to correspond to topologically
non-trivial surgery operations of the genus $g$ surface. \cite{barkeshli2012} These mathematical
operations correspond to cutting the space along a non-contractible loop, rotating by $2\pi$, and gluing back together.

\bf{Proposed Experiments} \rm

\it 1. Quantized zero-bias quasiparticle conductance \rm

The $\sqrt{m}$ topological degeneracy associated with a genon, {\it i.e.} an end point of the TLJ,
corresponds to a localized parafermion zero mode\cite{barkeshli2012a,barkeshli2012,clarke2012,lindner2012,cheng2012,vaezi2012,fendley2012},
which generalizes the usual Majorana zero modes in topological superconductors.  
Zero mode operators $\alpha_i$ can be defined at each genon, which satisfy the 'parafermion' algebra
$\alpha_i\alpha_j=\alpha_j\alpha_ie^{2\pi i/m},~i<j$ with a proper ordering of all genons in the system.
Physically, the operator $\alpha_i$ is a quasi-particle tunneling operator between the two layers at the
genon point, which commutes with the Hamiltonian. We refer to Ref. \cite{barkeshli2012} for more
rigorous definition of the zero mode operators.

\begin{figure}
\centerline{
\includegraphics[width=3.5in]{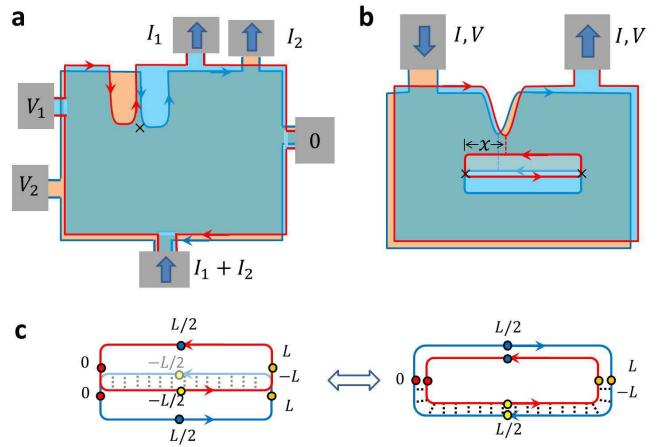}
}
\caption{\label{zeroBias}Experimental setups to measure zero bias interlayer quasiparticle conductance.
(a) The "Andreev reflection" process (see text) in which the exciton current $I_1-I_2$ is reversed across TLJ. 
(b) Use the edge states of the sample to measure the local quasiparticle density of states, which detect the parafermion zero modes at the ends of the TLJ.(c) Edge theory picture of TLJ in (b). The coordinate $\tilde{x}$ increase from $-L$ to $+L$ following the direction indicated by orange, yellow, red and blue dots. The second layer (blue) is flipped in the right panel.
Dotted lines between counterpropagating edge states indicate backscattering, generating an energy gap.
}
\end{figure}

In the case of the Majorana zero modes localized at the ends of a 1D topological superconductor, one of its
key signatures is a quantized zero bias conductance peak with conductance $G=\frac{2e^2}{h}$ when a normal
metal lead is coupled to the Majorana zero mode \cite{law2009,alicea2012review,fidkowski2012}. This is due to the perfect Andreev
reflection in the presence of the Majorana mode. In our system, the analog of an electron in topological superconductor is a quasi-particle-quasi-hole pair, {\it i.e.}, a ``topological exciton" with charge $(1/m,-1/m)$. Here the two numbers represent the charge of the exciton in the two layers. When such an exciton moves across the TLJ, the two layers are exchanged, such that its charge becomes $(-1/m,1/m)$; a topological exciton will become its anti-particle when going across a TLJ, just like an electron being transformed into a hole in a superconductor through an Andreev process. The topological exciton has a perfect ``Andreev" reflection at presence of the parafermion zero mode, which can be probed experimentally (Fig. \ref{zeroBias} (a) ). By separately contacting the two layers, the edge states can carry different currents, and the relative current $I_r=I_1-I_2$ is the exciton current. If we switch off the TLJ (by turning off the gate voltages), the two layers are decoupled and the voltage in each layer is proportional to its current. Therefore $dI_r/dV_r=\frac{1}{m}\frac{e^2}h$. With the TLJ the two layers are exchanged when the edge states meet the TLJ, so that the two voltages are also reversed, and $dI_r/dV_r=-\frac{1}{m}\frac{e^2}h$. The cases with and without TLJ correspond to perfect Andreev and perfect normal scattering of the exciton.

To verify that the zero bias peak conductance observed from the perfect ``Andreev" reflection experiment is really from a local parafermion zero mode, we can further consider the experimental setting in Fig. \ref{zeroBias} (b), which is an analog of using a scanning tunneling microscope (STM) to probe the Majorana zero mode. By contacting source and drain to the two different layers and bringing the edge states close to the TLJ, the edge states carry the exciton current $I_r$, and play the role of an STM tip. It is well-known that the electron tunneling from an STM tip leads to a differential conductance proportional to the local electron density of states. By analogy, the tunneling conductance in our system measures the local density of states of the exciton, related to $G_{(1,-1)} (x,\omega) = \langle V^\dagger_{(1,-1)}(x,\omega) V_{(1,-1)}(x,-\omega) \rangle$.
$V_{1,-1}(x,t) = e^{i \phi_{L1}(x,t) - i  \phi_{L2}(x,t)}$ annihilates an exciton in the upper edge of the TLJ, or equivalently, tunnels a
$1/m$ quasiparticle from layer $2$ to layer $1$. We find $G_{(1,-1)}$ has the following exponential behavior:
$G_{(1,-1)}(x,\omega) = f(x) e^{-x/\xi} \delta(\omega)$,
where $\xi \sim \hbar v/g$ is the correlation length of the gapped region (with $g$ the gap opened at the TLJ),
and $f(x)$ is a power-law function of $x$. The exponential localization is remarkable, since
part of the edge states are still gapless.
To understand this, consider the edge theory description of the TLJ in Fig. \ref{zeroBias} (b).
In Eq. (\ref{H0}), (\ref{Ht}),
we used separate chiral boson fields $\phi_{L(R)1},~\phi_{L(R)2}$ to
describe the two edges from each layer. Here, for convenience we
describe the whole edge of the first (second) layer by a single chiral boson $\phi_{1(2)}(x)$. Therefore
$\phi_{1,2}(x)$ each lives on a circle with perimeter $2L$ (with $L$ the length of the TLJ). The edge of
each layer is a chiral Luttinger liquid, and it is convenient to group the two chiral Luttinger liquids
into a non-chiral Luttinger liquid. For this purpose we define the spatial coordinate $\tilde{x}\in[-L,L]$
of the two edges with opposite chiralities (Fig. \ref{zeroBias} (c)). $\tilde{x}$ increases
along the counter-clockwise direction for layer $1$ and clockwise direction for layer $2$. With such
a choice, $\phi_1$ and $\phi_2$ has opposite chirality (defined with respect to $\tilde{x}$), and the
vertical back-scattering induced at the TLJ occurs locally between $\phi_1(\tilde{x})$ and $\phi_2(\tilde{x})$.
The Luttinger liquid Lagrangian can be written as
\begin{align}
L = & \int_{-L}^L d\tilde{x} [(\partial_t \varphi)^2 + v (\partial_{\tilde{x}} \varphi)^2
+ g \theta(-\tilde{x}) \cos( m \varphi ) ].
\end{align}
with $\varphi=\phi_1+\phi_2$ and where $\theta(x)$ is a step function. The tunneling induces a gap in the
region $ \tilde{x}\in[-L,0]$, and the region $\tilde{x}\in[0,L]$ remains gapless. 
In this representation, the $m$ topological degenerate ground states are correctly given by the $m$ minima of the $\cos(m\varphi)$ term, with $\varphi$ pinned to the values $2\pi n/m,~n=0,1,...,m-1$. Importantly, in this coordinate the exciton annihilation operator becomes nonlocal: $V_{1,-1}(x) = e^{i \phi_1(\tilde{x}) - i \phi_2(-\tilde{x})}$. 
Due to the non-local commutation relation of chiral boson fields, 
applying this operator to the ground state causes a domain wall in $\varphi$ in the region $-L < \tilde{x} < 0$. Away from the
defects at $x = 0$ and $x = -L \sim L$, such a domain wall costs an energy of order the gap $g$.
Therefore, the correlations of this operator must decay exponentially, away from the positions of the
defects. At the positions of the defects, this operator keeps the system in the ground state manifold because it does not create
a domain wall in $\varphi$ in the gapped region. Instead, it simply yields a unitary transformation among the ground states by globally flipping the value of $\varphi$ in the gapped region. From this analysis we see that the zero bias peak in the exciton local density of state is indeed probing the topologically degenerate states.

\it 2. Long-time current-noise cross correlation \rm

Since the genons are non-abelian defects, it should be possible to detect them through interferometry. Compared to the interferometry measurements
proposed for detecting non-Abelian anyons in non-abelian FQH states\cite{bonderson2006,stern2006,willett2009}, 
an advantage of our current setup is that the non-abelian defects (genons) are imposed externally
rather than being dynamical quasiparticle excitations, so that their position and number can be easily controlled.

\begin{figure}
\centerline{
\includegraphics[width=3.5in]{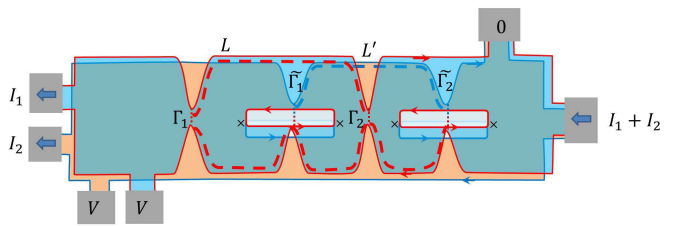}
}
\caption{\label{interferometer}QPC interferometer setup for detecting the non-Abelian nature of genons. $\Gamma_{1,2}$ are the two QPC's in the first layer, and $\tilde{\Gamma}_{1,2}$ are the two inter-layer QPC's across the TLJs. $L$ and $L'$ indicated by thick dashed lines are the two interference loops. 
}
\end{figure}

Our setup (Fig. \ref{interferometer})  consists of two TLJs and four quantum point contacts
(QPCs), created by either a top gate or a bottom gate. One pair of QPCs ($\Gamma_{1,2}$ in Fig. \ref{interferometer}) allows for weak quasiparticle tunneling across the normal region of the fluid, between the edges of the top layer. The other pair ($\tilde{\Gamma}_{1,2}$) allows for weak quasiparticle tunneling across the
TLJs, from one edge of the top layer to the other edge of the bottom layer. It should be noticed that there are only two nontrivial interference loops formed by the four QPCs, the first one ($L$ in Fig. \ref{interferometer}) in the top layer including $\Gamma_1$ and $\Gamma_2$, and the second one ($L'$) going through both layers including $\tilde{\Gamma}_1$ and $\tilde{\Gamma}_2$. Other interference paths such as the one enclosing $\Gamma_1$ and $\tilde{\Gamma}_1$ does not form a closed loop since the quasiparticle cannot return to the same layer. The interference in these two loops $L,L'$ measure the electric charge in the loops, which measure the topological qubit state of the genons $1,2$ and $2,3$ respectively. As a key consequence of the non-Abelian nature of genons, the topological qubit degrees of freedom are non-local, and the topological charge carried by genons $1,2$ does not commute with that of $2,3$. Therefore non-local correlation between interference patterns in these two loops are induced due to the uncertainty principle. Alternatively, the non-commutativity of the two quasiparticle paths $L$ and $L'$ can also be understood from the geometrical picture shown in Fig. \ref{doubleStaircase}, with the measured topological charge corresponding to $\hat{W}(a)$, $\hat{W}(b)$.

To be more precise, the interference pattern can be observed by applying a voltage difference $V$ between the left and right-moving edges in both layers,
and measuring the currents $I_1, ~I_2$ (Fig. \ref{interferometer}).
We find that the interlayer current noise cross-correlation
$S_{12}(t) \equiv \frac12 \langle \{I_{1}(t), I_{2}(0) \} \rangle - \langle I_{1}(t) \rangle \langle I_{2}(0) \rangle$ has a dramatic feature:
\begin{align}
S_{12}(|t| \gg 1/T) \rightarrow constant \neq 0.
\end{align}
Such a long time noise correlation, at finite temperatures, is a striking direct consequence of the
non-Abelian nature of the genons, which has never been proposed or observed before in an Abelian state.
Intuitively, this is understood as follows. The current $I_{1}(t)$ depends on the quasiparticle interference between the two QPCs $\Gamma_{1,2}$.
When a quasiparticle leaves the top layer through the QPC $\tilde{\Gamma}_1$, its fractional statistics
cause a phase shift in the interference of loop $L$ and therefore changes the value of $I_{1}(t)$. Such a
phase shift is permanent until the next quasi-particle tunneling occurs at $\tilde{\Gamma}_1$. The
converse is also true, that the quasiparticle tunneling at $\Gamma_2$ causes a permanent phase shift
in $I_{2}(t)$. Therefore, the fluctuation of the currents $I_{1}(t)$ and $I_{2}(0)$ have a long-time
correlation for both $t>0$ and $t<0$. It should be noticed that only having three QPC's
$\Gamma_1,\tilde{\Gamma}_1,\Gamma_2$ \footnote{The interferometer with only three QPCs
$\Gamma_1,\tilde{\Gamma}_1,\Gamma_2$ is topologically equivalent to a setup proposed in
\cite{kane2003},  as a way to measure fractional statistics.} can already induce a long-time correlation of $S_{12}$,
but only for $t>0$.

We present in the Supplemental Materials an explicit calculation of $S_{12}$ based on the edge theory;
the result, for $|t| \gg 1/T$ and to lowest order in the tunneling strength at the QPC's, takes the form:
\begin{align}
S_{12}(t) = &T^{\frac{4}{m} - 2} [ |\Gamma_1^2 \tilde{\Gamma}_1 \tilde{\Gamma}_2| \theta(-t) S_{12}^{(1)} +
|\tilde{\Gamma}_1^2 \Gamma_1 \Gamma_2| \theta(t) S_{12}^{(2)}
\nonumber \\
&+|\Gamma_1 \Gamma_2 \tilde{\Gamma}_1 \tilde{\Gamma}_2| (S_{12}^{(3)} \theta(t) + S_{12}^{(4)} \theta(-t) ) ].
\end{align}
The dimensionless functions $S_{12}^{(i)}$, whose explicit form is given in the Supplemental Materials,
depend on the quasiparticle statistics (they are non-zero only when $m \neq 1$)
and decay with the dimensionless ratios $\hbar eV /mk_B T$, and  $k_B T x /\hbar v$,
which are optimally of order one, where $x$ is the distance scale between the QPCs.

On much longer time scales, thermally excited mobile quasiparticles can decohere the state of the qubit, which will wash out
the noise signature considered here. However this time scale is expected to be exponentially large in $1/T$, (\it ie \rm $t \propto e^{\Delta/T}$),
because the quasiparticle density is exponentially small in $1/T$.

\bf{Discussion} \rm

A longer term goal is to implement unitary gates on the topological qubit by braiding genons.
This can be realized with further sets of gates to tune interactions
between various zero modes localized at the genons.\cite{alicea2010b,barkeshli2012,clarke2012,lindner2012}
The computational power of genons in generic Abelian states is stronger than that provided by
Majorana zero modes \cite{clarke2012}. The setup proposed here can be generalized to realize universal TQC
in non-abelian states that by themselves would not be universal for TQC \cite{barkeshli2012}. Examples include
two layers of the non-abelian phase of Kitaev's honeycomb model \cite{kitaev2006,barkeshli2012}, or the odd-denominator
Bonderson-Slingerland FQH states \cite{bonderson2008,bondersonpc}.

\it Acknowledgments \rm-- We would like to thank David Goldhaber-Gordon, Woo-Won Kang, Charlie Marcus and Hari Manoharan for helpful conversations. This work was supported by the Simons Foundation (MB) and David and Lucile Packard foundation (XLQ).


\newpage

\appendix

\newpage

\begin{widetext}

\section{Supplemental Materials}

\section{I. Parafermion Zero Modes}

As mentioned in the main text, an important property of the TLJs is that their ends
localize exotic forms of zero modes, called parafermion zero modes, which generalize
the usual Majorana zero mode algebra. \cite{barkeshli2012} Here we will briefly recall
the definition and physical meaning of these zero mode operators, reviewing results from
\cite{barkeshli2012}.

Let us imagine aligning all TLJs, and cutting the whole system along the line, yielding the counterpropagating
edge states. Then, the normal regions of the fluid can be obtained by gluing the edges back
together with the electron tunneling terms $\sum_I (\Psi_{eLI}^\dagger \Psi_{eRI} + H.c.) \propto \sum_I \cos(m \phi_I)$,
while the twisted line junctions can be described by gluing the edges back together with the twisted tunneling
$\Psi_{eL1}^\dagger \Psi_{eR2} + \Psi_{eL2}^\dagger \Psi_{eR1} + H.c \propto \sum_I \cos(m \tilde{\phi}_1)$,
where we have defined $\phi_I = \phi_{LI} + \phi_{RI}$, and $\tilde{\phi}_I = \phi_{LI} + \phi_{R (I+1)\% 2}$.
Here, we have included also the tunneling term $\Psi_{eL2}^\dagger \Psi_{eR1} + H.c$, which is not present
in the experimental setup of Fig. \ref{expsetup}.

The ends of the TLJ can therefore be considered as domain walls between these two
different ways of gapping the counterpropagating edge states:
\begin{align}
\delta \mathcal{H}_{tun} = g\left\{ \begin{array}{ccc}
\sum_I \cos(m \phi_I) & \text{ if }  & x \in A_i\\
\sum_i \cos(m \tilde{\phi}_I) & \text{ if } & x \in B_i  \\
\end{array} \right.
\end{align}
where $A_i$ indicates the normal regions, while $B_i$ are the regions associated with the TLJs.

As shown in \cite{barkeshli2012}, we can define quasiparticle tunneling operators (see Fig. \ref{edgeFig})
$\alpha_{2i-1} = e^{i\phi_1(x_{A_i})} e^{-i \tilde{\phi}_1(x_{B_i})}$,
$\alpha_{2i} = e^{i \tilde{\phi}_2(x_{B_i})} e^{-i \phi_2(x_{A_{i+1}})}$,
which are zero modes of the edge Hamiltonian, $[\alpha_i, H_{edge}] = 0$,
and which satisfy a $Z_m$ 'parafermion' algebra: $\alpha_i \alpha_j = \alpha_j \alpha_i e^{2\pi i/m}$,
and $\alpha_i^m = 1$.
\begin{figure}
\includegraphics[width=3in]{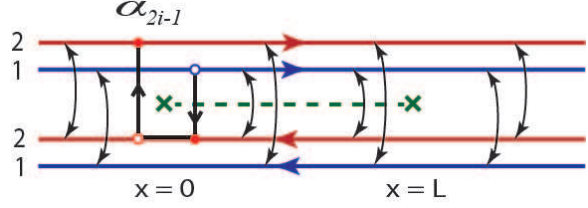}
\caption{\label{edgeFig} The defects (genons) can be viewed as domain walls between two different ways of generating an energy gap in the
counterpropagating edge states (double arrows indicate interedge electron tunneling). The zero mode operator $\alpha_{2i-1}$
corresponds to the quasiparticle hopping process shown in the figure, projected into the ground state subspace of
the edge theory.
}
\end{figure}

Other ways of creating domain walls between two different ways of generating an energy gap in
a single pair of counterpropagating chiral Luttinger liquids has been explored in
\cite{clarke2012,lindner2012,cheng2012}, where similar parafermion zero modes are obtained.

\section{II. Interdigitated gating}

In the experimental setup described in the main text, there are gapless edge states surrounding
the TLJs. These edge states do not affect the topological degeneracies, the exponential localization
of the zero modes, or the possibility of braiding the genons by tuning the interactions between
the parafermion zero modes. Nevertheless, it might be desirable to remove these gapless edge states,
while still preserving the topological degeneracy. This can be done simply by
alternating the offset in space (see Fig. \ref{zigzag}).
The tunneling term in the edge states can be modelled as follows:
\begin{align}
\delta \mathcal{H}_{tunn} = \alpha(x) \cos( m \tilde{\phi}_1) + \beta(x) \cos(m \tilde{\phi}_2) .
\end{align}
While $\alpha(x)$ and $\beta(x)$ are never both non-zero simultaneously, at long wavelengths only the average value
of $\alpha$ and $\beta$ is important. Therefore, both edge states can be gapped in the twisted way on average
in this setup.
\begin{figure}
\centerline{
\includegraphics[width=4in]{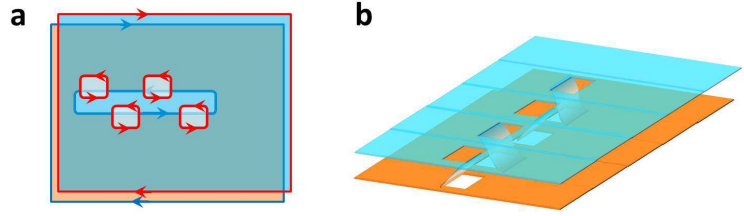}
}
\caption{\label{zigzag}Alternating the offset of the top and bottom gate will allow us to generate an energy gap for
all edge states. At long wavelengths, both twisted tunneling terms, $\sum_I \cos ( m \tilde{\phi}_I)$ are simultaneously
present.
}
\end{figure}

\section{III. Interlayer Current Noise Cross-Correlation }

\section{A. Summary and Results}

Here, will present the derivation of the current noise cross-correlation for the interferometer summarized in the main text.
We consider a tunneling Hamiltonian with four QPCs \cite{chamon1995,chamon1997}:
\begin{align}
\label{Htunn}
H_{tun}(t) &= \sum_{i=1}^2 [H_i(t) + \tilde{H}_i(t)],
\nonumber \\
H_i(t) &= v_i \mathcal{O}^+(x_i) + v_i^*  \mathcal{O}^-(x_i),
\nonumber \\
\tilde{H}_i(t) &= \tilde{v}_i \tilde{\mathcal{O}}^+(\tilde{x}_i) + \tilde{v}_i^* \tilde{\mathcal{O}}^-(\tilde{x}_i),
\end{align}
where
\begin{align}
v_i = \Gamma_i e^{-i \omega_1 t}, \;\;\; \tilde{v}_i = \tilde{\Gamma}_i e^{-i \omega_2 t}.
\end{align}
$\Gamma_i, \tilde{\Gamma}_i$ are the tunneling matrix elements for the different QPCs, and
$\omega_i = e^* V_i$, where $V_i$ are the potential differences across the QPCs and $e^* = e/m$ is the elementary fractional charge of the quasiparticles.
We note that in the setup of Fig. \ref{interferometer}, to lowest order in the tunneling matrix elements, the voltage across each QPC is
$V_i = V$. For what follows, it will be useful to define:
\begin{align}
x_{12} \equiv x_1 - x_2, \;\;\; \tilde{x}_{12} = \tilde{x}_1 - \tilde{x}_2.
\end{align}

The operator $\mathcal{O}^\pm$ is the quasiparticle tunneling operator across QPCs enclosing the normal regions of the fluid:
\begin{align}
\mathcal{O}^\pm(x_i) = \kappa_i^\pm e^{\pm i(\phi_{1L} - \phi_{1R}  ) }(x_i),
\end{align}
where $\kappa_i$ are Klein factors that ensure that the quasiparticle tunneling operators at different points in space commute with each other, as required.
The operators $\tilde{\mathcal{O}}^\pm$ are quasiparticle tunneling operators across the twisted line junctions (TLJs):
\begin{align}
\tilde{\mathcal{O}}^\pm(\tilde{x}_i) = \tilde{\kappa}_i^\pm e^{\pm i(\phi_{1L} - \phi_{2R}  ) }(\tilde{x}_i),
\end{align}
where $\tilde{\kappa}_i^\pm$ are also Klein factors. The commutation relations of $\kappa_i^\pm$ and $\tilde{\kappa}_i^\pm$
can be determined from the boson commutation relations:
\begin{align}
[\phi_{LI}(x),\phi_{LI}(x')] &= i \frac{\pi}{m} sgn(x - x')
\nonumber \\
[\phi_{RI}(x),\phi_{RI}(x')] &= -i \frac{\pi}{m} sgn(x - x'),
\end{align}
from which we determine that $[\kappa_1^r, \kappa_2^s] = 0$, $[\tilde{\kappa}_1^r, \tilde{\kappa}_2^s] = 0$, and
\begin{align}
\kappa_i^r \tilde{\kappa}_j^s = \tilde{\kappa}_j^s \kappa_i^r e^{i \theta_{ij}^{rs}},
\end{align}
where $\theta_{ij}^{rs} = rs \pi sgn(x_i - \tilde{x}_j)/m$.

In the interferometer discussed in the main text, which probes the non-abelian
character of the genons, we require $x_1 < \tilde{x}_1 < x_2 < \tilde{x}_2$. However it will also be helpful to compare this with
the case $x_1 < x_2 < \tilde{x}_1 < \tilde{x}_2$, which does not probe the non-commutativity of the two non-commuting, non-contractible
quasiparticle paths.

Now, we let $I_{b1}$ and $I_{b2}$ be the current that is backscattered through the normal regions and the TLJs, respectively:
\begin{align}
I_{bJ}(t) = \frac{d Q_{RJ}}{dt} = -i [Q_{RJ}, H_{tun}(t)],
\end{align}
where $Q_{RI} = \frac{1}{2\pi} \int dx \partial_x \phi_{RI}$ is the total charge on the right edge in the $I$th layer.
Thus:
\begin{align}
I_{b1}(t) &= \frac{i}{m}\sum_{i=1}^2[\Gamma_i e^{-i e^* V_1 t} \mathcal{O}^+(x_i) - H.c.] = \frac{i}{m} \sum_{i=1}^2 [v_i \mathcal{O}^+(x_i) - v_i^* \mathcal{O}^-(x_i)]
= \frac{i}{m} \sum_{i=1}^2 \sum_{\epsilon_i = \pm} \epsilon_i v_i^{\epsilon_i} \mathcal{O}^{\epsilon_i}(x_i)
\nonumber \\
I_{b2}(t) &= \frac{i}{m}\sum_{i=1}^2 [\tilde{\Gamma}_i e^{-i e^* V_2 t} \tilde{\mathcal{O}}^+(\tilde{x}_i) - H.c.]
=  \frac{i}{m}\sum_{i=1}^2 [\tilde{v}_i \tilde{\mathcal{O}}^+(\tilde{x}_i) - \tilde{v}_i^* \tilde{\mathcal{O}}^-(\tilde{x}_i)]
= \frac{i}{m} \sum_{i=1}^2 \sum_{\epsilon_i = \pm} \epsilon_i \tilde{v}_i^{\epsilon_i} \tilde{\mathcal{O}}^{\epsilon_i}(\tilde{x}_i)
\end{align}

In the setup of Fig. \ref{interferometer}, we find that:
\begin{align}
I_1 &= \frac{1}{m} \frac{e^2}{h} V + I_{b1},
\nonumber \\
I_2 &= \frac{1}{m} \frac{e^2}{h} V + I_{b2} + I_{g2},
\end{align}
where $I_1$ and $I_2$ are the currents flowing through the device in the first and second layer (see Fig. \ref{interferometer}).
In order to understand $I_{g2}$, observe that current can be backscattered through the TLJs in
two different ways. The first way is through quasiparticle tunneling across the TLJ. The second way
is if the outer edge that follows the constriction of the QPC overlaps the gapless edge states that
flow around the TLJ. In this case, there can be interlayer electron tunneling between the top and bottom
outer edges of the sample, through the edge states surrounding the TLJ. This latter contribution is what
we call $I_{g2}$. Therefore:
\begin{align}
S_{12}(t) &\equiv \frac12 \langle I_1(t) I_2(0) \rangle - \langle I_1(t) \rangle \langle I_2(0) \rangle
\nonumber \\
&= \frac12 \langle I_{b1}(t) (I_{b2}(0) + I_{g2}(0)) \rangle - \langle I_1(t) \rangle \langle (I_2(0) + I_{g2}(0)) \rangle
\end{align}
We will be interested in the limit $|t| \gg 1/T$, in which case it will become clear that $I_{g2}$, which involves
only electron tunneling, will not have long-time cross-correlations between the quasiparticle tunneling current
$I_{b1}$.

Therefore, we will compute the following interlayer noise cross-correlation:
\begin{align}
S_{12}(t) = \frac12 \langle \{I_{b1}(t), I_{b2}(0) \}\rangle - \langle I_{b1}(t) \rangle \langle I_{b2}(0) \rangle .
\end{align}
Note that the expectation values are given by
\begin{align}
\label{corrFun}
\langle A(t) \rangle = \langle T_C A(t) e^{-i \int_C  H_{tun}(\tau) d\tau} \rangle_0,
\end{align}
where the $\langle \cdots \rangle_0$ indicates an expectation value in the unperturbed theory. $C$ indicates the Keldysh
contour, and $T_C$ is time-ordering in the Keldsyh contour. In our calculation, we will drop the subscript and assume all expectation
values are in the unperturbed theory.

Let us first present the result of the calculation. We find that to lowest oder in the QPC tunneling matrix elements,
\begin{align}
S_{12}(|t| \gg 1/T) = &\frac{(\pi T)^{4/m - 2}}{m^2} [ |\Gamma_1|^2 |\tilde{\Gamma}_1 \tilde{\Gamma}_2|
\theta(-t) S_{12}^{(1)}(\bar{\omega}_2,\bar{\omega}_1, \bar{\tilde{x}}_{12}, \tilde{\varphi}_{12} ) +
 |\tilde{\Gamma}_1|^2 |\Gamma_1 \Gamma_2| \theta(t) S_{12}^{(1)}(\bar{\omega}_1, \bar{\omega}_2,\bar{x}_{12}, \varphi_{12}) +
\nonumber \\
&|\Gamma_1 \Gamma_2 \tilde{\Gamma}_1 \tilde{\Gamma}_2| (S_{12}^{(2)}(\bar{\omega}_1, \bar{\omega}_2, \bar{x}_{12}, \bar{\tilde{x}}_{12}, \tilde{\varphi}_{12}, \varphi_{12}) \theta(t) +
S_{12}^{(2)} (\bar{\omega}_2, \bar{\omega}_1, \bar{\tilde{x}}_{12}, \bar{x}_{12}, \varphi_{12}, \tilde{\varphi}_{12}) \theta(-t) ) ].
\end{align}
where we have defined the dimensionless quantities
\begin{align}
\bar{\omega}_1 &\equiv \frac{\hbar}{\pi k_B} \frac{\omega_1}{T}, \;\;\; \bar{\omega}_2 \equiv \frac{\hbar}{\pi k_B} \frac{\omega_2}{T},
\nonumber \\
\bar{x}_{12} &\equiv \frac{\pi k_B}{\hbar v} T x_{12} , \;\;\; \bar{\tilde{x}}_{12} \equiv \frac{\pi k_B}{\hbar v} T \tilde{x}_{12}
\end{align}
and $\varphi_{12}$ and $\tilde{\varphi}_{12}$ are the phase differences between the tunneling matrix elements
of the QPCs: $\Gamma_1 \Gamma_2^* = |\Gamma_1 \Gamma_2| e^{i \varphi_{12}}$, and
$\tilde{\Gamma}_1 \tilde{\Gamma_2}^* = |\tilde{\Gamma}_1\tilde{\Gamma}_2| e^{i\tilde{\varphi}_{12}}$.
Note that $S_{12}^{(i)}$ depend on only dimensionless quantities.
\begin{align}
\label{corrResult}
S_{12}^{(1)}(\bar{\omega}_2, \bar{\omega}_1, x, \tilde{\varphi}_{12}) =&  4 c_r(\bar{\omega}_2,2 x)
[ (\cos(2\pi/m) - 1) \cos \tilde{\varphi}_{12} f_-(\bar{\omega}_1, 0)
- \sin \tilde{\varphi}_{12} \sin(2\pi/m) f_+(\bar{\omega}_1, 0)] ,
\nonumber \\
S_{12}^{(2)}(\bar{\omega}_1, \bar{\omega}_2, x, y, \tilde{\varphi}_{12}, \varphi_{12}) =&
8 c_r(\bar{\omega}_1, 2 x) \cos \tilde{\varphi}_{12} [ 2 (\cos(\pi/m) - 1)  \cos(\varphi_{12}) c_r(\bar{\omega}_2,2 y)
\nonumber \\
&- \sin(\pi/m) \sin(\varphi_{12})  f_+(\bar{\omega}_2, y)],
\end{align}
The functions $f_{\pm}(\bar{\omega}, \bar{x})$ are defined as
\begin{align}
f_{\pm}(\bar{\omega}, \bar{x}) \equiv
\int_{-\infty}^{\infty} e^{ i \bar{\omega} t} dt [e^{-\frac{i\pi}{2m} (sgn(t + \bar{x}) + sgn(t - \bar{x}) ) } \pm e^{\frac{i\pi}{2m} (sgn(t + \bar{x}) + sgn(t - \bar{x}) ) }]
\left( \frac{1}{\sinh |t + \bar{x}| \sinh |t - \bar{x}|} \right)^{1/m}
\end{align}
while $c_r(a,b)$ is defined as
\begin{align}
c_r(a,b) \equiv -2 \sin(\pi/m) \text{ Im}\left\{ e^{i ab/2} \int_0^{\infty} dt e^{i a t} \left(\frac{1}{\sinh[t] \sinh [t+ |b|]} \right)^{1/m} \right\}
\end{align}
\begin{figure}[b]
\centering
\subfigure[]{\includegraphics[width=2.2in]{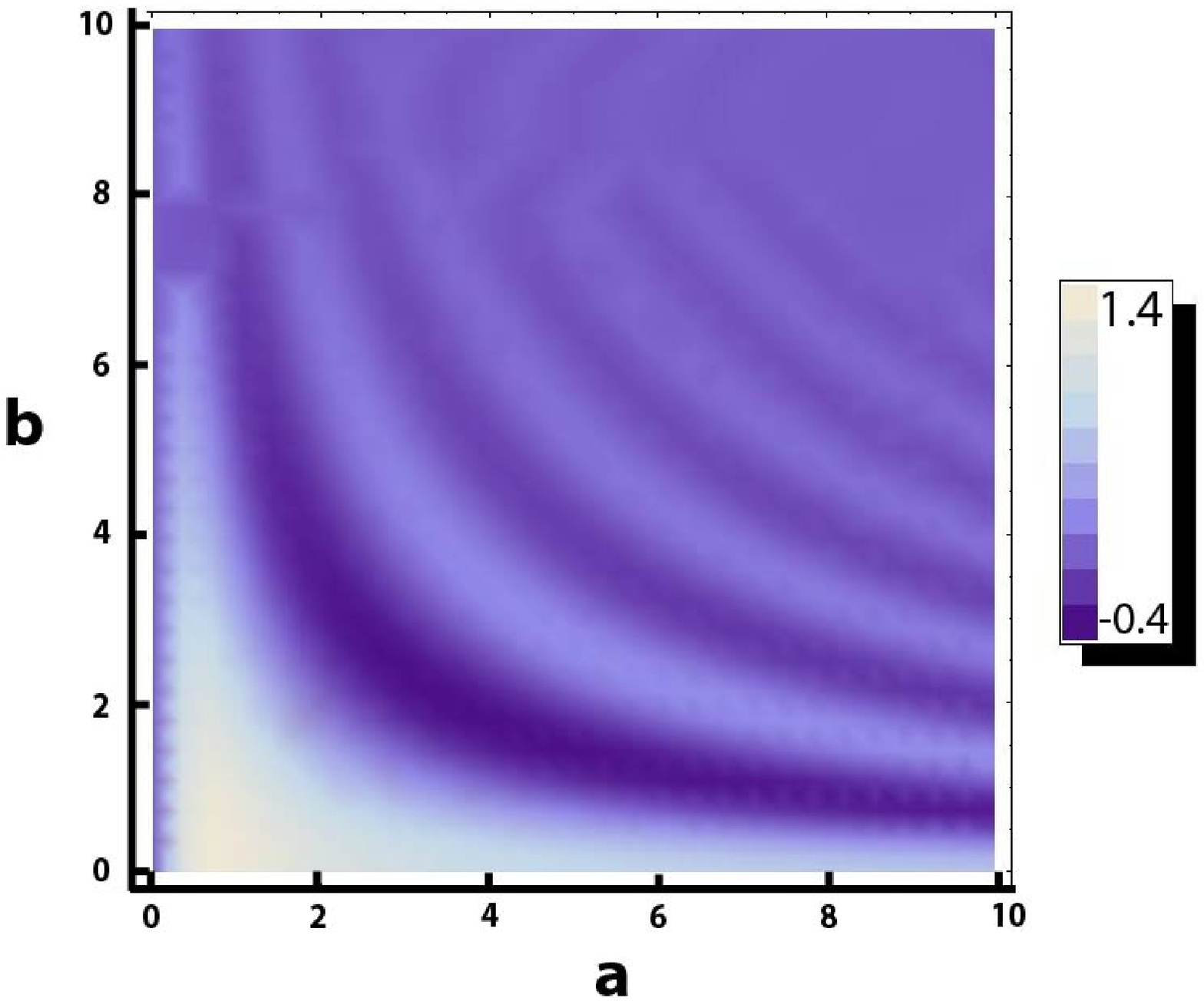}}
\subfigure[]{\includegraphics[width=2.3in]{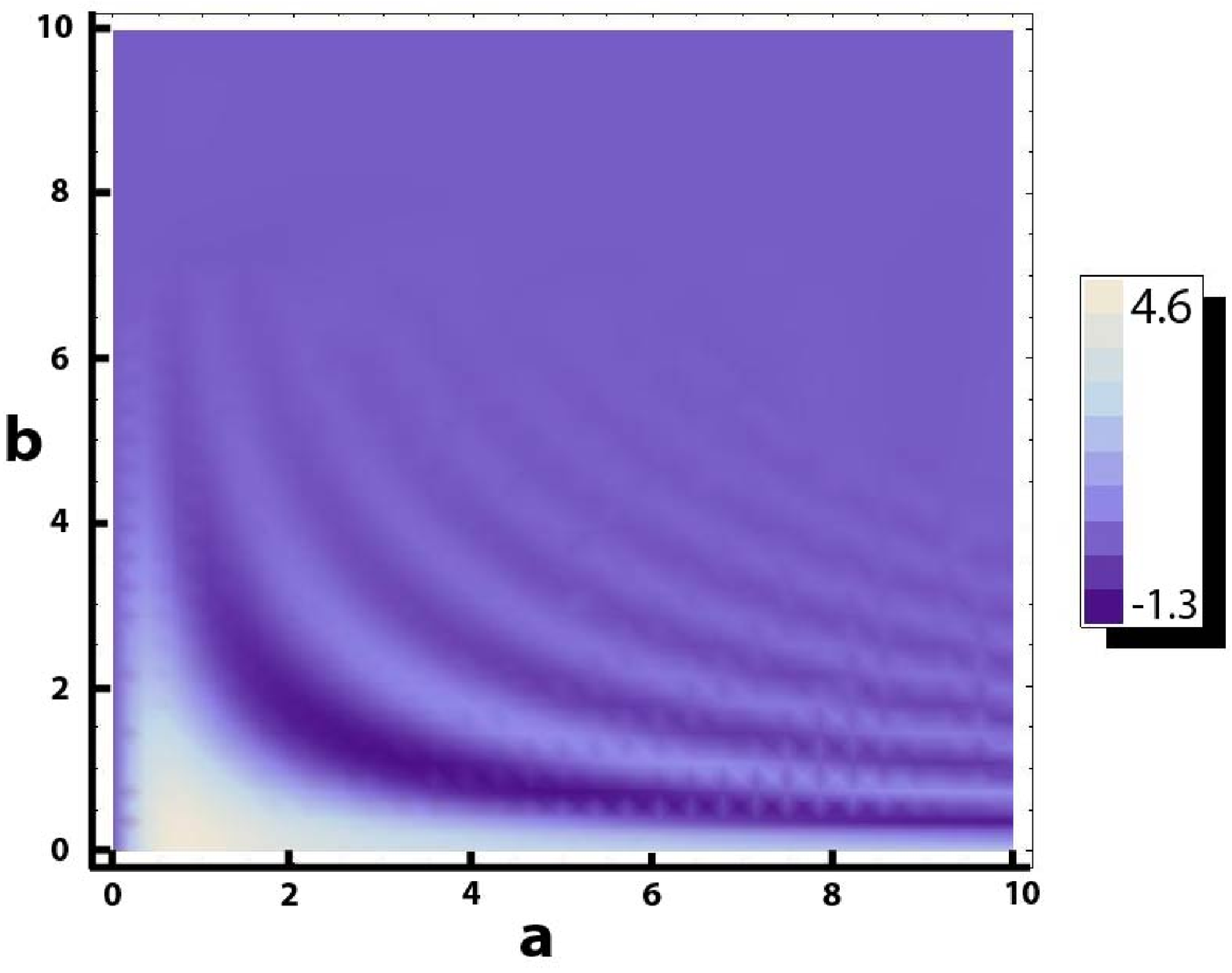}}
\subfigure[]{\includegraphics[width=2.2in]{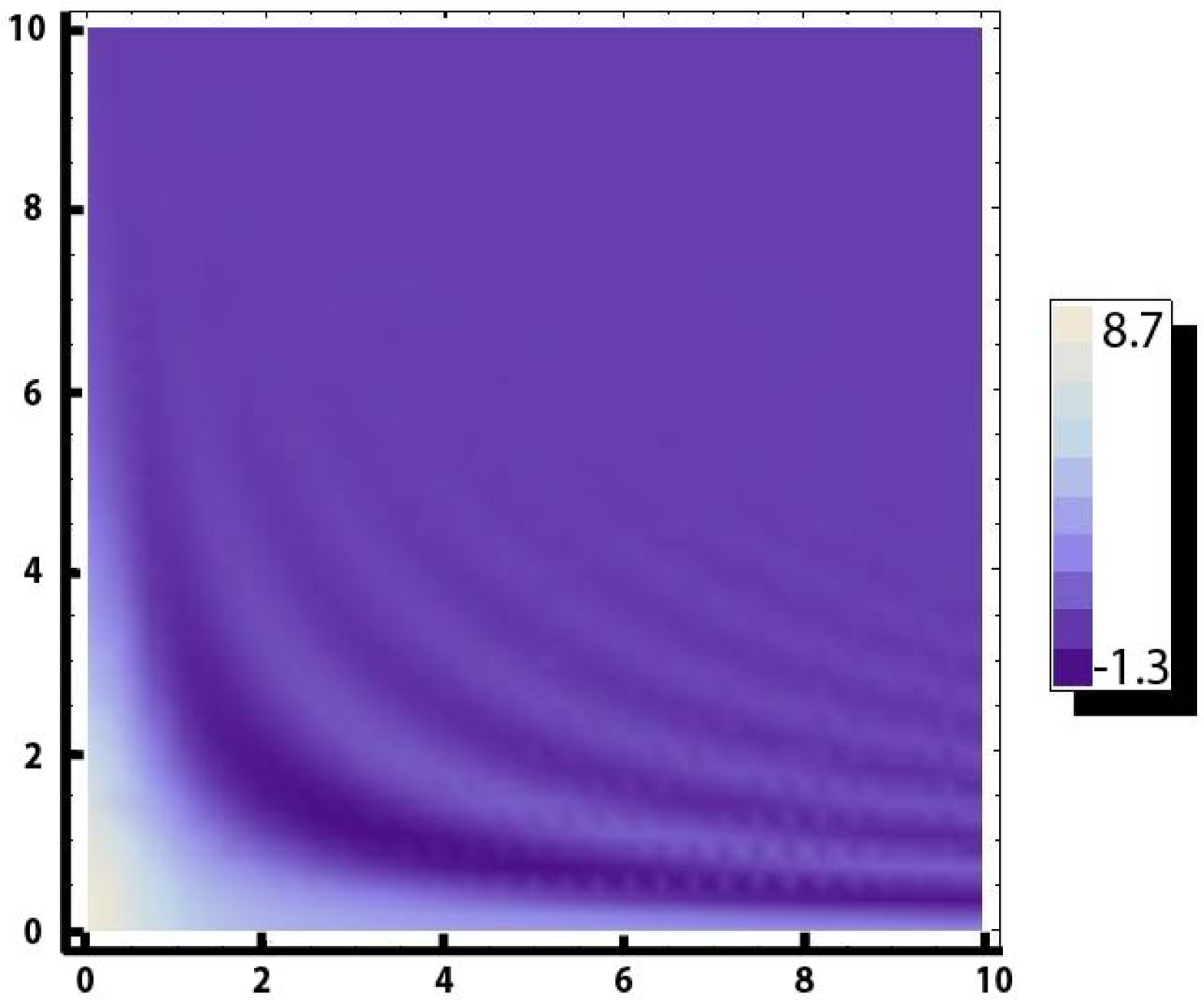}}
\caption{Density plots of (a) $-\frac{c_r(a,b)}{2 \sin(\pi/m)}$, (b) $f_-(a,b)$, and (c) $f_+(a,b)$, for $m = 3$.
\label{plotsF} }
\end{figure}
As a check, observe that when $m = 1$ and $m \rightarrow \infty$, the two non-contractible paths for quasiparticle propagation commute, and
therefore this noise cross-correlation should vanish at long times, which it indeed does.

In Fig. \ref{plotsF}, we plot the functions $f_{\pm}(a,b)$ and $-c_r(a,b)/\sin(\pi/m)$. In Fig \ref{plotsS}, we plot
$S_{12}^{(i)}(a,b,a,b,0,0)$, for $i =1,2$; that is, for the case where the voltage differences between
all QPCs is the same, $x_{12} = \tilde{x}_{12}$, and for zero phase differences $\varphi_{12}$ and $\tilde{\varphi}_{12}$.
We see that when $a$ and $b$ are of order one, $S_{12}^{(i)}$ reach their maximum value, and decay quickly as
$a$ and $b$ increase. From this, we see that the noise cross-correlation will require the dimensionless
parameters $\bar{\omega}_i$, $\bar{x}_{12}$, and $\tilde{\bar{x}}_{12}$ to all be of order one.
If we assume $T \approx 50-100 mK$, and $v \approx 10^5 m/s$, we see that this requires $x_{12}$ and $\tilde{x}_{12}$
be on the order of several $\mu m$, which is reasonable for realistic device dimensions.

\begin{figure}[ht]
\centering
\subfigure[]{\includegraphics[width=2.5in]{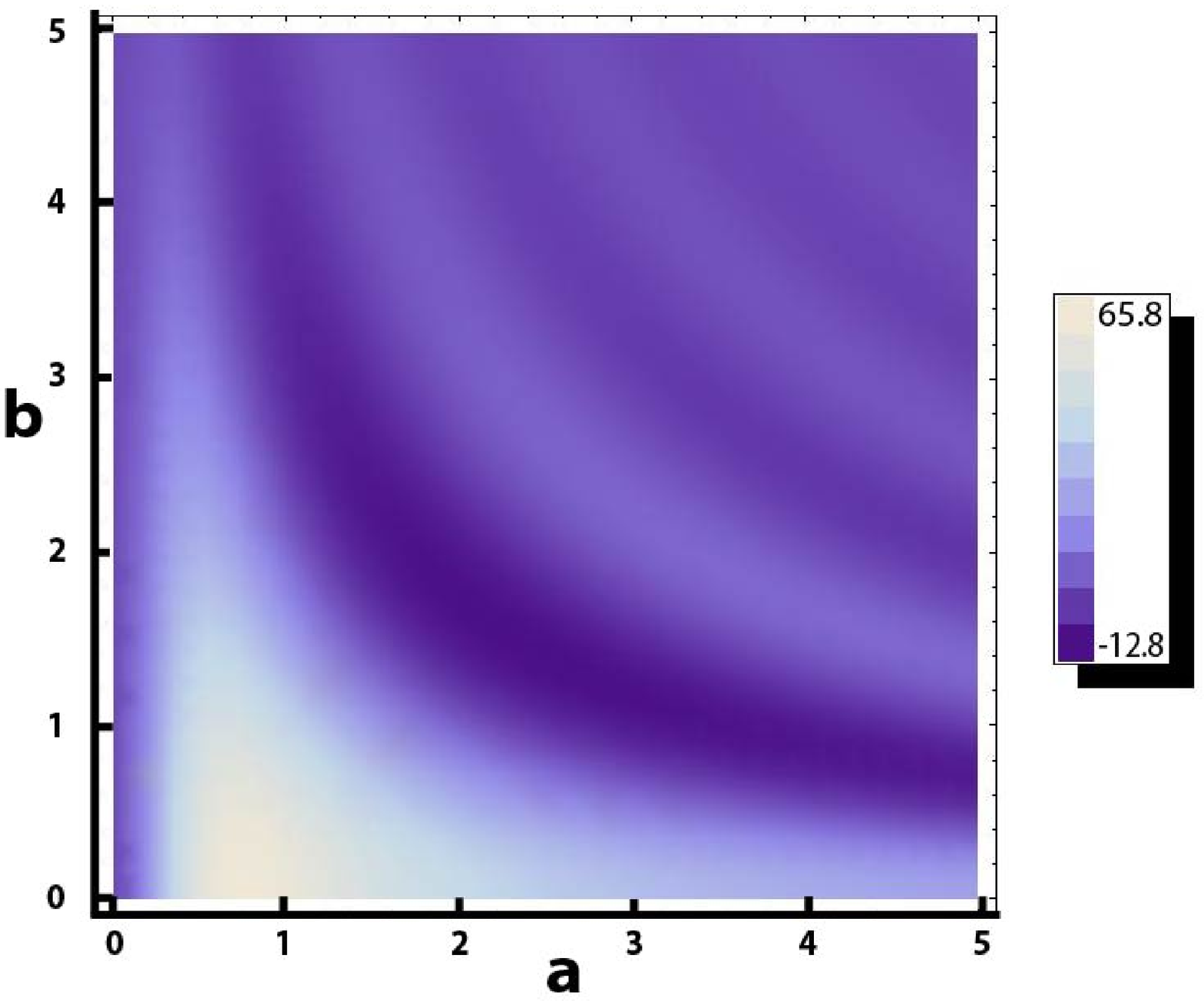}}
\subfigure[]{\includegraphics[width=2.5in]{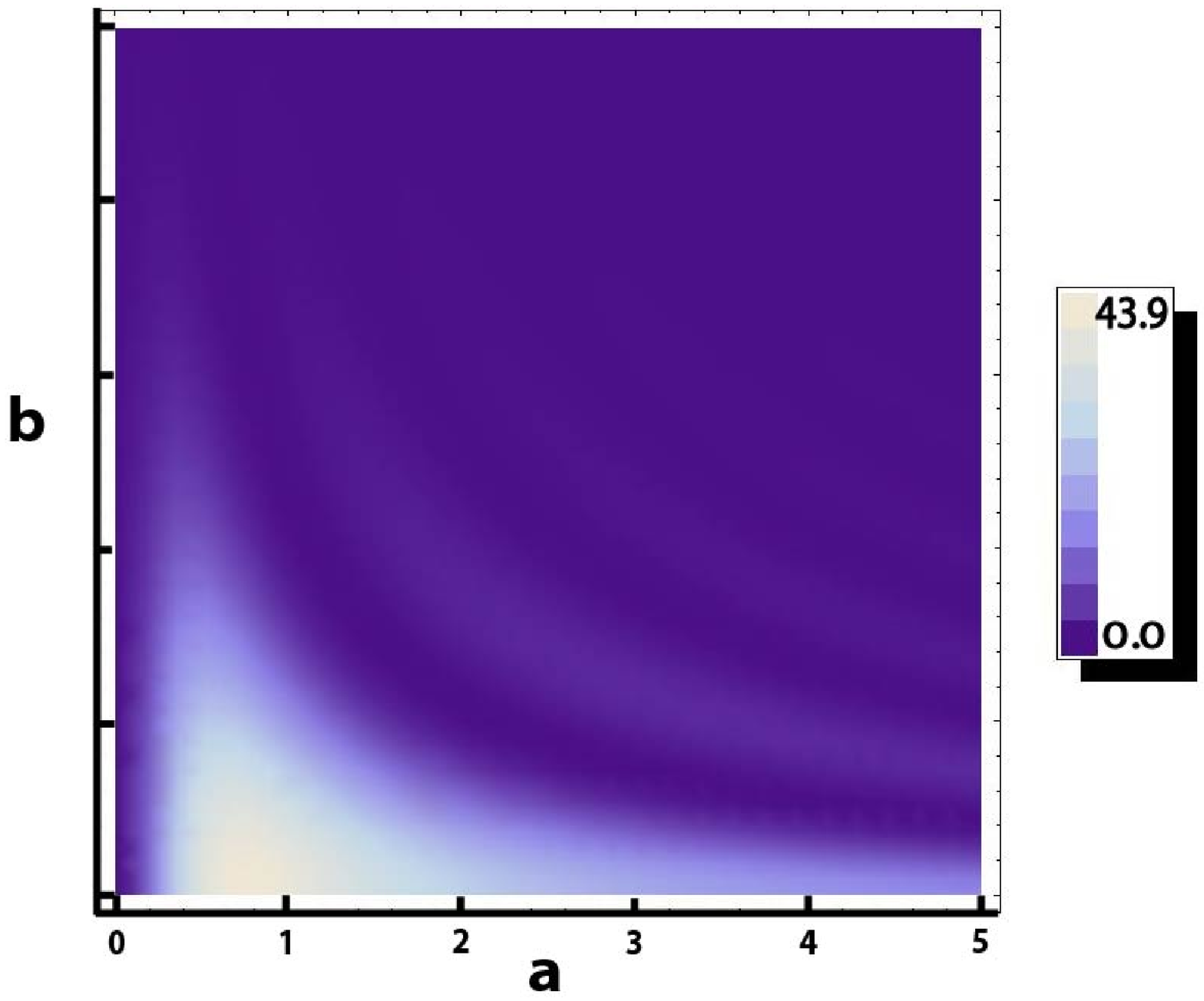}}
\caption{Density plots of (a) $S_{12}^{(1)}(a,b,a,b,0,0)$ and (b) $-S_{12}^{(2)}(a,b,a,b,0,0)$, for $m = 3$.
Both are zero when $a = b = 0$, reach a maximum for $a$ and $b$ of order one, and decay to zero with
increasing $a$ and $b$.
\label{plotsS} }
\end{figure}

\subsection{B. Klein Factors}

Here, we will determine the Keldysh-contour ordered correlation functions of the Klein factors, which are required for the calculation.
To do this, first we observe that the two point function for Klein factors that commute with each other should be one. Next, we observe that the correlation functions
of several Klein factors should factor into two-point functions. In order to obtain the two-point function of non-commuting Klein factors,
it is helpful to represent the Klein factors in terms of bosonic operators \cite{guyon2002}:

\begin{align}
\kappa_i^\pm = :e^{\pm i \sqrt{1/m} \theta_i}:,
\;\;\;\;
\tilde{\kappa}_i^\pm = :e^{\pm i \sqrt{1/m} \tilde{\theta}_i } :,
\end{align}
where $\theta_i, \tilde{\theta}_i$ are bosonic operators and for what follows we will leave the normal ordering
$:\cdots :$ implicit. Using the fact that
$:e^A : : e^B :  = :e^B : :e^A: e^{[A,B]}$
for cases where $[A,B]$ is a $c$-number, we find that the commutation relation of the Klein factors is reproduced when
\begin{align}
[\theta_i, \tilde{\theta_j}] = -i\pi sgn(x_i - \tilde{x}_j), \;\;\;\;\; [\theta_i, \theta_j ] = [\tilde{\theta_i}, \tilde{\theta}_j] = 0
\end{align}
Since $\theta_i$ and $\tilde{\theta}_j$ are conjugates, we can form a harmonic oscillator algebra:
\begin{align}
a_{ij} = \frac{1}{\sqrt{2\pi}} (\theta_i + i \tilde{\theta_j} ), \;\;\;\;\; a_{ij}^\dagger = \frac{1}{\sqrt{2\pi}} (\theta_i - i \tilde{\theta_j} ),
\end{align}
such that
\begin{align}
[a_{ij},a_{ij}^\dagger ] = - sgn(x_i - \tilde{x}_j), \;\;\;
\langle a_{ij} a_{ij}^\dagger \rangle = \frac{1 - sgn(x_i - \tilde{x}_j)}{2} .
\end{align}

We suppose that the Hamiltonian for these bosonic fields is zero, as they have no dynamics.
From this we conclude:
\begin{align}
\langle \theta_i(t) \theta_i(0) \rangle = \langle \tilde{\theta}_i(t) \tilde{\theta}_i(0) \rangle = \pi/2,
\;\;\;\;\;
\langle \theta_i(t) \tilde{\theta}_j(0) \rangle = - \langle \tilde{\theta}_j(t) \theta_i(0) \rangle = -\frac{i \pi}{2} sgn(x_i - \tilde{x}_j).
\end{align}
From this, we conclude that the Keldysh contour ordered correlation function is:
\begin{align}
g_{ij}^{\eta_1 \eta_2}(t) \equiv \langle T_C \theta_i (t^{\eta_1}) \tilde{\theta}_j(0^{\eta_2}) \rangle = -i \frac{\pi}{2} sgn(x_i - \tilde{x}_j)\chi_{\eta_1 \eta_2}(t),
\end{align}
where
\begin{align}
\chi_{\eta_1 \eta_2}(t) =   \frac{\eta_1 + \eta_2}{2} sgn(t) - \frac{\eta_1 - \eta_2}{2}.
\end{align}
Therefore:
\begin{align}
\langle T_C \kappa_i^{\epsilon}(t^{\eta_1}) \tilde{\kappa}^{\tilde{\epsilon}}_j(0^{\eta_2}) \rangle = e^{i \epsilon \tilde{\epsilon} \frac{\pi}{2m} \chi_{\eta_1 \eta_2}(t) sgn(x_i - \tilde{x}_j)}
\end{align}

For later reference, we will need the following correlation function:
\begin{align}
\langle T_C \kappa_{\alpha_1}^{\epsilon_1} (t^\eta) \kappa_{\beta_1}^{-\epsilon_1}(t^{\eta_1}) \tilde{\kappa}_{\alpha_2}^{\epsilon_2}(0^{-\eta}) \tilde{\kappa}_{\beta_2}^{-\epsilon_2}(0^{\eta_2}) \rangle
= \langle T_C e^{i \sqrt{1/m} \epsilon_1 \theta_{\alpha_1}(t^{\eta})} e^{-i \sqrt{1/m} \epsilon_1 \theta_{\beta_1}(t^{\eta_1})} e^{i \sqrt{1/m} \epsilon_2 \tilde{\theta}_{\alpha_2} (0^{-\eta})} e^{- \epsilon_2 i \sqrt{1/m} \tilde{\theta}_{\beta_2}(0^{\eta_2}))} \rangle
\end{align}
To calculate this, we use the fact that
$\langle T_C e^{A_1} \cdots e^{A_n} \rangle = e^{\sum_{i < j}^n \langle T_C  A_i A_j \rangle}$,
where in the Klein factor case, we only include correlation functions between non-commuting variables.
Therefore:
\begin{align}
\langle T_C \kappa_{\alpha_1}^{\epsilon_1} (t^\eta) \kappa_{\beta_1}^{-\epsilon_1}(t^{\eta_1}) \tilde{\kappa}_{\alpha_2}^{\epsilon_2}(0^{-\eta}) \tilde{\kappa}_{\beta_2}^{-\epsilon_2}(0^{\eta_2}) \rangle
&= e^{i \frac{\pi}{2m} \epsilon_1 \epsilon_2 (\chi_{\eta -\eta}(t) s_{\alpha_1 \alpha_2} - \chi_{\eta \eta_2}(t) s_{\alpha_1 \beta_2} - \chi_{\eta_1 -\eta}(t) s_{\beta_1 \alpha_2} + \chi_{\eta_1 \eta_2}(t) s_{\beta_1 \beta_2})}
\nonumber \\
&= \left\{ \begin{array}{cc}
e^{ -i \frac{\epsilon_1 \epsilon_2}{2m} \pi [ \eta (s_{\alpha_1 \alpha_2} - s_{\beta_1 \alpha_2}) + \eta_2 (s_{\alpha_1 \beta_2} - s_{\beta_1 \beta_2})]} & t > 0 \\
e^{ i \frac{\epsilon_1 \epsilon_2}{2m} \pi [ \eta (s_{\alpha_1 \beta_2} - s_{\alpha_1 \alpha_2}) + \eta_1 (s_{\beta_1 \alpha_2} - s_{\beta_1 \beta_2})]} & t < 0 \\
\end{array}\right.
\end{align}
where $s_{\alpha \beta} \equiv sgn( x_{\alpha} - \tilde{x}_\beta)$.
When $x_1 < x_2 < \tilde{x}_1 < \tilde{x}_2$, $s_{\alpha \beta} = -1$ for all $\alpha$, $\beta$, and therefore
\begin{align}
\langle T_C \kappa_{\alpha_1}^{\epsilon_1} (t^\eta) \kappa_{\beta_1}^{-\epsilon_1}(t^{\eta_1}) \tilde{\kappa}_{\alpha_2}^{\epsilon_2}(0^{-\eta}) \tilde{\kappa}_{\beta_2}^{-\epsilon_2}(0^{\eta_2}) \rangle
= 1.
\end{align}
When $x_1 < \tilde{x}_1 < x_2 < \tilde{x}_2$, which is the case of interest for our proposed interferometer, we instead have
\begin{align}
\langle T_C \kappa_{\alpha_1}^{\epsilon_1} (t^\eta) \kappa_{\beta_1}^{-\epsilon_1}(t^{\eta_1}) \tilde{\kappa}_{\alpha_2}^{\epsilon_2}(0^{-\eta}) \tilde{\kappa}_{\beta_2}^{-\epsilon_2}(0^{\eta_2}) \rangle
= \left\{ \begin{array}{ccc}
e^{ - i\frac{\epsilon_1 \epsilon_2}{m} (\alpha_1 - \beta_1) \pi [ \eta \delta_{\alpha_2 1}+ \eta_2 \delta_{\beta_2 1}]} & \text{ if }  & t > 0 \\
 e^{-i \frac{\epsilon_1 \epsilon_2}{m} (\alpha_2 - \beta_2)\pi [ -\eta \delta_{\alpha_1 2} + \eta_1 \delta_{\beta_1 2}]} & \text{ if } & t < 0  \\
\end{array} \right.
\end{align}

\subsection{C. Boson correlation functions}

We will need the following correlation functions:
\begin{align}
\label{bosonCor1}
G_{\eta \eta'}(x-x',t-t') \equiv \langle T_C e^{i \epsilon \phi(x,t^\eta)} e^{-i \epsilon \phi(x', t'^{\eta'} ) } \rangle = e^{\langle T_C \phi(x,t^\eta) \phi(x', t'^{\eta'}) \rangle}.
\end{align}
At finite temperatures, this is (setting the velocities of the left- and right- movers to unity):
\begin{align}
\label{bosonCor2}
G_{\eta_1 \eta_2}(x,t)=
\left(\frac{(\pi T)^2}{\sin[\pi T(\delta + i\chi_{\eta_1 \eta_2}(t) (t+x))] \sin [\pi T( \delta + i \chi_{\eta_1 \eta_2}(t) (t-x))]} \right)^{1/m},
\end{align}
where
\begin{align}
\chi_{\eta_1 \eta_2}(t) = \frac{(\eta_1 + \eta_2)}{2} sgn(t) - \frac{(\eta_1 - \eta_2)}{2}
\end{align}
Taking the limit $\delta \rightarrow 0^+$, we can replace the above by
\begin{align}
G_{\eta_1 \eta_2}(x,t)=
e^{-i \frac{\pi}{2m} \chi_{\eta_1 \eta_2}(t) ( sgn(t+x) + sgn(t-x) ) } \left(\frac{(\pi T)^2}{\sinh[\pi T|t+x|] \sinh [\pi T|t-x|]} \right)^{1/m}
\end{align}
The chiral correlator, involving only the left or right-moving modes (and re-inserting the velocities $v_{L/R}$ for the left/right movers), is:
\begin{align}
G^{L/R}_{\eta_1 \eta_2} (x, t) =\langle T_C e^{i \epsilon \phi_{L/R}(x,t^{\eta_1})} e^{-i \epsilon \phi_{L/R}(0, 0^{\eta_2})} \rangle =
e^{-i \frac{\pi}{2m} \chi_{\eta_1 \eta_2}(t)  sgn(|v|  t \pm x) } \left(\frac{(\pi T)}{\sinh[\pi T||v_{L/R}| t \pm x|]} \right)^{1/m}
\end{align}

\subsection{D. Details of calculation}

Here we describe in detail the calculation of $S_{12}(t)$, to lowest order in the tunneling matrix elements. We have (recall eq. (\ref{Htunn}), (\ref{corrFun}) ):
\begin{align}
\frac12 \langle \{I_{b1}(t), I_{b2}(0) \}\rangle &= \frac14\sum_{\eta = \pm} (-i)^2 \int_C d\tau_1 d\tau_2 \langle T_C I_{b1}(t^\eta) H_{tun}(\tau_1) I_{b2}(0^{-\eta}) H_{tun}(\tau_2) \rangle
\nonumber \\
&= - \frac12 \sum_{\eta,\eta_1, \eta_2 = \pm}  \int_{-\infty}^{\infty} \eta_1 \eta_2 d\tau_1^{\eta_1}  d\tau_2^{\eta_2} \langle T_C I_{b1}(t^\eta) H(\tau_1^{\eta_1}) I_{b2}(0^{-\eta}) \tilde{H}(\tau_2^{\eta_2}) \rangle,
\end{align}
Consider the correlation function:
\begin{align}
\langle T_C I_{b1}(t^\eta) H(\tau_1^{\eta_1}) I_{b2}(0^{-\eta}) \tilde{H}(\tau_2^{\eta_2}) \rangle =
&\frac{i^2}{m^2} \sum_{\alpha_1,\alpha_2,\beta_1,\beta_2} \sum_{\epsilon_{\alpha_1} \epsilon_{\beta_1} \epsilon_{\alpha_2} \epsilon_{\beta_2} }
v_{\alpha_1}^{\epsilon_{\alpha_1}} v_{\beta_1}^{\epsilon_{\beta_1}} v_{\alpha_2}^{\epsilon_{\alpha_2}} v_{\beta_2}^{\epsilon_{\beta_2}} \epsilon_{\alpha_1} \epsilon_{\alpha_2}
\nonumber \\
&\langle T_C \mathcal{O}^{\epsilon_{\alpha_1}}(x_{\alpha_1}, t^\eta) \mathcal{O}^{\epsilon_{\beta_1}}(x_{\beta_1}, \tau_1^{\eta_1})
\tilde{\mathcal{O}}^{\epsilon_{\alpha_2}}(\tilde{x}_{\alpha_2}, 0^{-\eta}) \tilde{\mathcal{O}}^{\epsilon_{\beta_2}}(\tilde{x}_{\beta_2}, \tau_2^{\eta_2}) \rangle
\end{align}
The above correlation function vanishes unless $\epsilon_{\alpha_1} = -\epsilon_{\beta_1}$ and $\epsilon_{\alpha_2} = - \epsilon_{\beta_2}$. Therefore:
\begin{align}
\langle T_C I_{b1}(t^\eta) H(\tau_1^{\eta_1}) I_{b2}(0^{-\eta}) \tilde{H}(\tau_2^{\eta_2}) \rangle =
&-\frac{1}{m^2}\sum_{\alpha_1,\alpha_2,\beta_1,\beta_2} \sum_{\epsilon_{1} \epsilon_{2} }
v_{\alpha_1}^{\epsilon_1} v_{\beta_1}^{-\epsilon_1} v_{\alpha_2}^{\epsilon_2} v_{\beta_2}^{-\epsilon_2} \epsilon_1 \epsilon_2
\langle T_C \kappa_{\alpha_1}^{\epsilon_1} (t^\eta) \kappa_{\beta_1}^{-\epsilon_1}(\tau_1^{\eta_1}) \tilde{\kappa}_{\alpha_2}^{\epsilon_2}(0^{-\eta}) \tilde{\kappa}_{\beta_2}^{-\epsilon_2}(\tau_2^{\eta_2}) \rangle
\nonumber \\
&\langle T_C e^{-i \epsilon_1 \phi_{1R}(x_{\alpha_1}, t^\eta)} e^{i \epsilon_1 \phi_{1R}( x_{\beta_1}, \tau_1^{\eta_1})} \rangle
\langle T_C e^{-i \epsilon_2 \phi_{2R} (\tilde{x}_{\alpha_2}, 0^{-\eta})} e^{i \epsilon_2 \phi_{2R}(\tilde{x}_{\beta_2},\tau_2^{\eta_2}) } \rangle
\nonumber \\
&\langle T_C e^{i \epsilon_1 \phi_{1L}(x_{\alpha_1}, t^\eta)} e^{-i \epsilon_1 \phi_{1L}(x_{\beta_1}, \tau_1^{\eta_1}) } e^{i \epsilon_2 \phi_{1L}(x_{\alpha_2}, 0^\eta)} e^{-i \epsilon_2 \phi_{1L}(\beta_2, \tau_2^{\eta_2})} \rangle
\end{align}
Recall that in our notation, the superscript $^-$ implies complex conjugation, not inverse.
\begin{align}
\langle T_C I_{b1}(t^\eta) H(\tau_1^{\eta_1}) I_{b2}(0^{-\eta}) \tilde{H}(\tau_2^{\eta_2}) \rangle =
&-\frac{1}{m^2}\sum_{\epsilon_1, \epsilon_2 = \pm} \sum_{\alpha_1, \beta_1, \alpha_2, \beta_2} \epsilon_1 \epsilon_2
\Gamma_{\alpha_1}^{\epsilon_1} \Gamma_{\beta_1}^{-\epsilon_1} \Gamma_{\alpha_2}^{\epsilon_2} \Gamma_{\beta_2}^{-\epsilon_2}
e^{i \epsilon_1 \omega_1 (t- \tau_1) + i \epsilon_2 \omega_2 (-\tau_2)}
\nonumber \\
&\langle T_C \kappa_{\alpha_1}^{\epsilon_1} (t^\eta) \kappa_{\beta_1}^{-\epsilon_1}(\tau_1^{\eta_1}) \tilde{\kappa}_{\alpha_2}^{\epsilon_2}(0^{-\eta}) \tilde{\kappa}_{\beta_2}^{-\epsilon_2}(\tau_2^{\eta_2}) \rangle
\nonumber \\
& G^{R,v_{1R}}_{-\eta \eta_2}(\tilde{x}_{\alpha_2} - \tilde{x}_{\beta_2}, -\tau_2) G^{R,v_{2R}}_{\eta \eta_1} (x_{\alpha_1} - x_{\beta_1}, t - \tau_1)
\nonumber \\
&\langle T_C e^{i \epsilon_1 \phi_{L1}(x_{\alpha_1}, t^\eta)} e^{-i \epsilon_1 \phi_{L1}(x_{\beta_1}, \tau_1^{\eta_1}) }
e^{i \epsilon_2 \phi_{L1}(\tilde{x}_{\alpha_2}, 0^{-\eta})} e^{-i \epsilon_2 \phi_{L1}(\tilde{x}_{\beta_2}, \tau_2^{\eta_2})} \rangle
\end{align}
This gives:
\begin{align}
\langle T_C I_{b1}(t^\eta) H(\tau_1^{\eta_1}) I_{b2}(0^{-\eta}) \tilde{H}(\tau_2^{\eta_2}) \rangle =
&-\frac{1}{m^2}\sum_{\epsilon_1, \epsilon_2 = \pm} \sum_{\alpha_1, \beta_1, \alpha_2, \beta_2} \epsilon_1 \epsilon_2
\Gamma_{\alpha_1}^{\epsilon_1} \Gamma_{\beta_1}^{-\epsilon_1} \Gamma_{\alpha_2}^{\epsilon_2} \Gamma_{\beta_2}^{-\epsilon_2}
e^{i \epsilon_1 \omega_1 (t- \tau_1) + i \epsilon_2 \omega_2 (-\tau_2)}
\nonumber \\
&\langle T_C \kappa_{\alpha_1}^{\epsilon_1} (t^\eta) \kappa_{\beta_1}^{-\epsilon_1}(\tau_1^{\eta_1})
\tilde{\kappa}_{\alpha_2}^{\epsilon_2}(0^{-\eta}) \tilde{\kappa}_{\beta_2}^{-\epsilon_2}(\tau_2^{\eta_2}) \rangle
\nonumber \\
& G^{R, v_{2R}}_{-\eta \eta_2}(\tilde{x}_{\alpha_2} - \tilde{x}_{\beta_2}, -\tau_2) G^{L, v_{1L}}_{-\eta \eta_2}(\tilde{x}_{\alpha_2} - \tilde{x}_{\beta_2}, -\tau_2)
\nonumber \\
&G^{R; v_{1R}}_{\eta \eta_1} (x_{\alpha_1} - x_{\beta_1}, t - \tau_1) G^{L; v_{1L}}_{\eta \eta_1} (x_{\alpha_1} - x_{\beta_1}, t - \tau_1)
\nonumber \\
&\left(
\frac{G^{L, v_{1L}}_{\eta \eta_2}(x_{\alpha_1} - \tilde{x}_{\beta_2}, t - \tau_2) G^{L, v_{1L}}_{\eta_1 -\eta}(x_{\beta_1} - \tilde{x}_{\alpha_2}, \tau_1)}
{G^{L, v_{1L}}_{\eta -\eta}(x_{\alpha_1} - \tilde{x}_{\alpha_2}, t) G^{L, v_{1L}}_{\eta_1 \eta_2} ( x_{\beta_1} - \tilde{x}_{\beta_2}, \tau_1 - \tau_2) }
\right)^{\epsilon_1 \epsilon_2}
\end{align}

The Green's functions $G^{L/R}_{\eta \eta'}(x, t)$ decay exponentially in $t$; therefore,  the integral will only be
dominated by regions where $\tau_1 \approx t$ and $\tau_2 \approx 0$. Furthermore, we are interested in
the long time behavior, where $t$ is large. In this limit, the ratio in parantheses above will go to one:
\begin{align}
\left(
\frac{G^{L, v_{1L}}_{\eta \eta_2}(x_{\alpha_1} - \tilde{x}_{\beta_2}, t - \tau_2) G^{L, v_{1L}}_{\eta_1 -\eta}(x_{\beta_1} - \tilde{x}_{\alpha_2}, \tau_1)}
{G^{L, v_{1L}}_{\eta -\eta}(x_{\alpha_1} - \tilde{x}_{\alpha_2}, t) G^{L, v_{1L}}_{\eta_1 \eta_2} ( x_{\beta_1} - \tilde{x}_{\beta_2}, \tau_1 - \tau_2) }
\right) \rightarrow 1
\end{align}
when $t \approx \tau_1 \rightarrow \infty$ and $|t| \gg |\tau_2|$, $t \gg |x_{\alpha} - \tilde{x}_\beta|$.
Therefore:
\begin{align}
\langle T_C I_{b1}(t^\eta) H(\tau_1^{\eta_1}) I_{b2}(0^{-\eta}) \tilde{H}(\tau_2^{\eta_2}) \rangle \approx
-\frac{1}{m^2}\sum_{\epsilon_1, \epsilon_2 = \pm} \sum_{\alpha_1, \beta_1, \alpha_2, \beta_2} \epsilon_1 \epsilon_2
&\Gamma_{\alpha_1}^{\epsilon_1} \Gamma_{\beta_1}^{-\epsilon_1} \Gamma_{\alpha_2}^{\epsilon_2} \Gamma_{\beta_2}^{-\epsilon_2}
e^{i \epsilon_1 \omega_1 (t- \tau_1) + i \epsilon_2 \omega_2 (-\tau_2)}
\nonumber \\
&
\langle T_C \kappa_{\alpha_1}^{\epsilon_1} (t^\eta) \kappa_{\beta_1}^{-\epsilon_1}(\tau_1^{\eta_1}) \tilde{\kappa}_{\alpha_2}^{\epsilon_2}(0^{-\eta}) \tilde{\kappa}_{\beta_2}^{-\epsilon_2}(\tau_2^{\eta_2}) \rangle
\nonumber \\
& G^{v_{2R};v_{1L}}_{-\eta \eta_2}(\tilde{x}_{\alpha_2} - \tilde{x}_{\beta_2}, -\tau_2) G^{v_{1R};v_{1L}}_{\eta \eta_1}( x_{\alpha_1} - x_{\beta_1}, t - \tau_1)
\end{align}

Therefore, so far we have, for large $t$,
\begin{align}
S_{12}(t) =  &\frac12 \frac{1}{m^2} \sum_{\eta,\eta_1, \eta_2 \epsilon_1, \epsilon_2 = \pm} \sum_{\alpha_1, \beta_1, \alpha_2, \beta_2}
\epsilon_1 \epsilon_2 \Gamma_{\alpha_1}^{\epsilon_1} \Gamma_{\beta_1}^{-\epsilon_1} \Gamma_{\alpha_2}^{\epsilon_2} \Gamma_{\beta_2}^{-\epsilon_2}
\nonumber \\
&[\langle T_C \kappa_{\alpha_1}^{\epsilon_1} (t^\eta) \kappa_{\beta_1}^{-\epsilon_1}(t^{\eta_1}) \tilde{\kappa}_{\alpha_2}^{\epsilon_2}(0^{-\eta}) \tilde{\kappa}_{\beta_2}^{-\epsilon_2}(0^{\eta_2}) \rangle -
\langle T_C \kappa_{\alpha_1}^{\epsilon_1} (t^\eta) \kappa_{\beta_1}^{-\epsilon_1}(t^{\eta_1}) \rangle \langle T_C \tilde{\kappa}_{\alpha_2}^{\epsilon_2}(0^{-\eta}) \tilde{\kappa}_{\beta_2}^{-\epsilon_2}(0^{\eta_2}) \rangle
]
\nonumber \\
&\eta_1 \eta_2 \int_{-\infty}^{\infty} e^{ i \epsilon_2 \omega_2 (-\tau_2)} G^{v_{2R};v_{1L}}_{-\eta \eta_2}(\tilde{x}_{\alpha_2} - \tilde{x}_{\beta_2}, -\tau_2) d\tau_2
\int_{-\infty}^{\infty} d\tau_1 e^{i \epsilon_1 \omega_1 (- \tau_1)} G^{v_{1R};v_{1L}}_{\eta \eta_1}( x_{\alpha_1} - x_{\beta_1}, t - \tau_1)
\end{align}

In the case where $x_1 < x_2 < \tilde{x}_1 < \tilde{x}_2$, using the formula for the Klein factor expression found earlier, we find that
the difference of Klein factor correlation functions vanishes, so that $S_{12}(t) \rightarrow 0$ for $|t| \gg 1/T$.

Now let us focus on the case $x_1 < \tilde{x}_1 <  x_2 < \tilde{x}_2$.
In this case, for $t > 0$,
\begin{align}
S_{12}(t>0) =  &\frac12 \frac{1}{m^2}\sum_{\eta,\eta_1, \eta_2 \epsilon_1, \epsilon_2 = \pm} \sum_{\alpha_1, \beta_1, \alpha_2, \beta_2}  \epsilon_1 \epsilon_2
\Gamma_{\alpha_1}^{\epsilon_1} \Gamma_{\beta_1}^{-\epsilon_1} \tilde{\Gamma}_{\alpha_2}^{\epsilon_2} \tilde{\Gamma}_{\beta_2}^{-\epsilon_2}
[e^{ -i\frac{\epsilon_1 \epsilon_2}{m} (\alpha_1 - \beta_1) \pi [ \eta \delta_{\alpha_2 1}+ \eta_2 \delta_{\beta_2 1}]} - 1]
\nonumber \\
&\eta_1 \eta_2 \int_{-\infty}^{\infty} e^{ i \epsilon_2 \omega_2 (-\tau_2)} G^{v_{2R};v_{1L}}_{-\eta \eta_2}(\tilde{x}_{\alpha_2} - \tilde{x}_{\beta_2}, -\tau_2)  d\tau_2
\int_{-\infty}^{\infty} d\tau_1 e^{i \epsilon_1 \omega_1 (- \tau_1)} G^{v_{1R};v_{1L}}_{\eta \eta_1}( x_{\alpha_1} - x_{\beta_1}, t - \tau_1)
\end{align}

In what follows, let us assume that the edge velocities are the same on the left and right edges, and in the two layers, which we set to unity:
$v_{1L} = v_{1R} = v_{2R}  = 1$.

Now let us define:
\begin{align}
A_{\eta_1 \eta_2}^{\omega}(x) 
&= \int_{-\infty}^{\infty} e^{ i  \omega \tau} d\tau G_{\eta_1 \eta_2}(x,\tau)
\nonumber \\
&= \int_{-\infty}^\infty dt e^{i \omega t} e^{-i \frac{\pi}{2m} \chi_{\eta_1 \eta_2}(t) ( sgn(t+x) + sgn(t-x) ) }
\left(\frac{(\pi T)^2}{\sinh[\pi T|t+x|] \sinh [\pi T|t-x|]} \right)^{1/m}
\end{align}
Note that this satisfies
\begin{align}
A_{+-}^\omega (x) = A_{-+}^{-\omega} (x), \;\;\; A_{\eta_1 \eta_2}^{\omega}(x) = A_{\eta_1 \eta_2}^{\omega}(-x), \;\;\; A_{++}^\omega(x) + A_{--}^\omega(x) = A_{+-}^\omega(x) + A_{-+}^\omega(x).
\end{align}

Then we can write the noise as a simple expression:
\begin{align}
S_{12}(t > 0) = \frac12 \frac{1}{m^2} \sum_{\eta,\eta_1, \eta_2 \epsilon_1, \epsilon_2 = \pm} \sum_{\alpha_1, \beta_1, \alpha_2, \beta_2}  \epsilon_1 \epsilon_2 \eta_1 \eta_2
&| \Gamma_{\alpha_1} \Gamma_{\beta_1} \tilde{\Gamma}_{\alpha_2} \tilde{\Gamma}_{\beta_2} |
e^{i \epsilon_1 \varphi_{\alpha_1 \beta_1} + i \epsilon_2 \tilde{\varphi}_{\alpha_2 \beta_2}}
\nonumber \\
&[e^{ -i\frac{\epsilon_1 \epsilon_2}{m} (\alpha_1 - \beta_1) \pi [ \eta \delta_{\alpha_2 1}+ \eta_2 \delta_{\beta_2 1}]} - 1]
A_{\eta \eta_1}^{\epsilon_1 \omega_1}(x_{\alpha_1 \beta_1}) A_{-\eta \eta_2}^{\epsilon_2 \omega_2}(\tilde{x}_{\alpha_2 \beta_2})
\end{align}
and
\begin{align}
S_{12}(t < 0) = \frac12 \frac{1}{m^2} \sum_{\eta,\eta_1, \eta_2 \epsilon_1, \epsilon_2 = \pm} \sum_{\alpha_1, \beta_1, \alpha_2, \beta_2}  \epsilon_1 \epsilon_2 \eta_1 \eta_2
&| \Gamma_{\alpha_1} \Gamma_{\beta_1} \tilde{\Gamma}_{\alpha_2} \tilde{\Gamma}_{\beta_2} |
e^{i \epsilon_1 \varphi_{\alpha_1 \beta_1} + i \epsilon_2 \tilde{\varphi}_{\alpha_2 \beta_2}}
\nonumber \\
&[e^{-i \frac{\epsilon_1 \epsilon_2}{m} (\alpha_2 - \beta_2)\pi [ -\eta \delta_{\alpha_1 2} + \eta_1 \delta_{\beta_1 2}]} - 1]
A_{\eta \eta_1}^{\epsilon_1 \omega_1}(x_{\alpha_1 \beta_1}) A_{-\eta \eta_2}^{\epsilon_2 \omega_2}(\tilde{x}_{\alpha_2 \beta_2})
\end{align}

Now let us try to make more explicit the above expressions. We start with the $t > 0$ case.
Summing over $\alpha_1$ and $\beta_1$ gives:
\begin{align}
\label{S12tpos}
S_{12}(t>0) =  &\frac12 \frac{1}{m^2}| \Gamma_1 \Gamma_2| \sum_{\eta,\eta_1, \eta_2 \epsilon_1, \epsilon_2 = \pm} \sum_{\alpha_2, \beta_2} \epsilon_1 \epsilon_2
|\tilde{\Gamma}_{\alpha_2} \tilde{\Gamma}_{\beta_2} |
e^{i \epsilon_1 \varphi_{12} + i \epsilon_2 \tilde{\varphi}_{\alpha_2 \beta_2}}
(e^{i \epsilon_1 \epsilon_2 \frac{\pi}{m} (\eta \delta_{\alpha_2 1}+ \eta_2 \delta_{\beta_2 1})} - 1)
\nonumber \\
&\eta_1 \eta_2 A^{\epsilon_2 \omega_2}_{-\eta \eta_2} (\tilde{x}_{12}) (A^{\epsilon_1 \omega_1}_{\eta \eta_1}(x_{12}) - A^{-\epsilon_1 \omega_1}_{\eta \eta_1}(x_{12}))
\end{align}
For $S_{12}(t<0)$, we perform the sum over $\alpha_2$, $\beta_2$ to get:
\begin{align}
\label{S12tneg}
S_{12}(t < 0) = \frac12 \frac{1}{m^2}\sum_{\eta,\eta_1, \eta_2 \epsilon_1, \epsilon_2 = \pm} \sum_{\alpha_1, \beta_1}  \epsilon_1 \epsilon_2 \eta_1 \eta_2
&| \Gamma_{\alpha_1} \Gamma_{\beta_1} \tilde{\Gamma}_{1} \tilde{\Gamma}_{2} |
e^{i \epsilon_1 \varphi_{\alpha_1 \beta_1} + i \epsilon_2 \tilde{\varphi}_{12}}
\nonumber \\
&[e^{i \frac{\epsilon_1 \epsilon_2}{m} \pi [ -\eta \delta_{\alpha_1 2} + \eta_1 \delta_{\beta_1 2}]} - 1]
A_{\eta \eta_1}^{\epsilon_1 \omega_1}(x_{\alpha_1 \beta_1}) (A_{-\eta \eta_2}^{\epsilon_2 \omega_2}(\tilde{x}_{12}) - A_{-\eta \eta_2}^{-\epsilon_2 \omega_2}(\tilde{x}_{12}))
\end{align}

This can be written as the sum of two terms:
\begin{align}
S_{12}(t>0) = S_{12}^1(t>0) + S_{12}^2(t>0),
\end{align}
where $S_{12}^1(t>0)$ contains the terms where $\alpha_2 \neq \beta_2$ in the above sum. 
$S_{12}^2(t >0 )$ contains terms in the sum (\ref{S12tpos}) where $\alpha_2 = \beta_2$.
Similarly, we write
\begin{align}
S_{12}(t < 0 ) = S_{12}^1(t<0) + S_{12}^2(t<0),
\end{align}
where $S_{12}^1(t < 0)$ contains terms in the sum (\ref{S12tneg}) where $\alpha_1 \neq \beta_1$, while $S_{12}^2(t < 0)$ contains terms in the
sum where $\alpha_1 = \beta_1$.
After simplifying, we find
\begin{align}
S_{12}^1(t > 0) &= \frac{(\pi T)^{4/m - 2}}{m^2} |\Gamma_1 \Gamma_2 \tilde{\Gamma}_1 \tilde{\Gamma}_2|
S_{12}^{(2)}(\bar{\omega_1}, \bar{\omega_2}, \pi Tx_{12}, \pi T \tilde{x}_{12}, \tilde{\varphi}_{12}, \varphi_{12}) , \;\;\;
\nonumber \\
S_{12}^2(t > 0) &= \frac{(\pi T)^{4/m - 2}}{m^2}|\tilde{\Gamma}_1|^2 |\Gamma_1 \Gamma_2| S_{12}^{(1)}(\bar{\omega_1}, \bar{\omega_2}, \pi T x_{12}, \varphi_{12}),
\nonumber \\
S_{12}^1(t < 0) &=  \frac{(\pi T)^{4/m - 2}}{m^2}|\Gamma_1 \Gamma_2 \tilde{\Gamma}_1 \tilde{\Gamma}_2|S_{12}^{(2)}
(\bar{\omega}_2, \bar{\omega}_1, \pi T \tilde{x}_{12}, \pi T x_{12}, \varphi_{12}, \tilde{\varphi}_{12}) , \;\;\;
\nonumber \\
S_{12}^2(t < 0) &= \frac{(\pi T)^{4/m - 2}}{m^2}|\Gamma_1|^2 |\tilde{\Gamma}_1 \tilde{\Gamma}_2|S_{12}^{(1)}(\bar{\omega}_2,\bar{\omega}_1, \pi T \tilde{x}_{12}, \tilde{\varphi}_{12} ),
\end{align}
where $S_{12}^{(i)}$ are presented in (\ref{corrResult}).

\end{widetext}

\end{document}